\documentclass[%
 amsmath,amssymb,
 aps,
]{revtex4-2}

\usepackage{graphicx}
\usepackage{dcolumn}
\usepackage{bm}

\usepackage{physics}
\usepackage{mathrsfs}
\usepackage{textgreek}
\usepackage{subcaption}
\usepackage{wrapfig}

\usepackage{graphicx}
\usepackage[algo2e,linesnumbered,ruled,vlined]{algorithm2e}
\usepackage{xcolor}

\newcommand{\rt}{\tilde{\rho}}
\newcommand{\PSF}{\mathrm{PSF}}
\newcommand{\OTF}{\mathrm{OTF}}
\newcommand{\SNR}{\mathrm{SNR}}
\newcommand{\vsig}{\vec{\bf y}}
\newcommand{\bsig}{\bf y}
\newcommand{\kt}{{\bf k}_{\perp}}
\newcommand{\brh}{\pmb{\rho}}

\newcommand{\vn}{\vec{\bf n}}
\newcommand{\vc}{\vec{\pmb{c}}}
\newcommand{\vv}{\vec{\bf v}}
\newcommand{\vgr}{\vec{\bf g}}
\newcommand{\mA}{{\bf A}}
\newcommand{\mAadj}{{\bf A}^\dagger}

\newcommand{\iDFT}{{\bf F}^\dagger}
\newcommand{\DFT}{{\bf F}}
\newcommand{\OTFvec}{\vec{\pmb{{\Psi}}}}
\newcommand{\PSFvec}{\vec{\pmb{{\psi}}}}

\newcommand{\vN}{\vec{\bf N}}
\newcommand{\vC}{\vec{\bf C}}
\newcommand{\vY}{\vec{\bf Y}}
\newcommand{\bpsi}{{\pmb{\psi}}}
\newcommand{\bPsi}{{\pmb{\Psi}}}
\newcommand{\bc}{{\pmb{c}}}
\newcommand{\bv}{{\bf v}}
\newcommand{\bpad}{\mathrm{\bf pad}}
\newcommand{\fl}{{\bf \mathrm{\rm flatten}}}
\newcommand{\diag}{{\bf \mathrm{\rm diag}}}

\usepackage{amsmath}
\newmuskip\pFqmuskip

\newcommand*\pFq[6][8]{%
  \begingroup 
  \pFqmuskip=#1mu\relax
  \mathchardef\normalcomma=\mathcode`,
  \mathcode`\,=\string"8000
  \begingroup\lccode`\~=`\,
  \lowercase{\endgroup\let~}\pFqcomma
  {}_{#2}\tilde{F}_{#3}{\left[\genfrac..{0pt}{}{#4}{#5};#6\right]}%
  \endgroup
}
\newcommand{\pFqcomma}{{\normalcomma}\mskip\pFqmuskip}

\begin{document}

\preprint{APS/123-QED}

\title{Super resolution computational saturated absorption microscopy}
\thanks{Super-resolution Deconvolution Imaging (SDI)}%

\author{Gabe Murray}
 \altaffiliation[]{Department of  Physics, Colorado State University, Fort Collins, CO 80523, USA}
\author{Jeff Field}
 \altaffiliation[]{Department of Electrical and Computer Engineering, Colorado State University, Fort Collins, CO 80523, USA}
\author{Patrick Stockton}
 \altaffiliation[]{Department of Electrical and Computer Engineering, Colorado State University, Fort Collins, CO 80523, USA}
\author{Ali Pezeshki}
 \altaffiliation[]{Department of Electrical and Computer Engineering, Colorado State University, Fort Collins, CO 80523, USA}
\author{Jeff Squier}
 \altaffiliation[]{Department of Physics, Colorado School of Mines, Golden, CO 80401, USA}
\author{Randy Bartels}%
 \altaffiliation[]{Electrical and Computer Engineering Department}
  \email{randy.bartels@colostate.edu.}
\affiliation{ 
Colorado State University, Fort Collins, CO 80523 USA
}%

\date{\today}


\begin{abstract}
Imaging beyond the diffraction limit barrier has attracted wide attention due to the ability to resolve image features that were previously hidden. Of the various super-resolution microscopy techniques available, a particularly simple method called saturated excitation microscopy (SAX) requires only a simple modification of a laser scanning microscope where the illumination beam power is sinusoidally modulated and driven into saturation. SAX images are extracted from harmonics of the modulation frequency and exhibit improved spatial resolution. Unfortunately, this elegant strategy is hindered by the incursion of shot noise that prevents high resolution imaging in many realistic scenarios. Here, we demonstrate a new technique for super resolution imaging that we call computational saturated absorption (CSA) in which a joint deconvolution is applied to a set of images with diversity in spatial frequency support among the point spread functions used in the image formation with saturated laser scanning fluorescence microscope. CSA microscopy allows access to the high spatial frequency diversity in a set of saturated effective point spread functions, while avoiding image degradation from shot noise.  
\end{abstract}

\maketitle


\section{\label{sec:level1}Introduction:\protect\\}

Optical imaging is a pervasive tool for observing the world due to the fact that optical radiation can non-destructively interrogate complex objects to perform a wide array of useful tasks. Conventional imaging strategies are limited in their ability to resolve fine spatial features due to what was previously viewed as a fundamental limitation on the ability to resolve features significantly smaller than the optical wavelength \cite{DiffLimit}. This limitation stems from the fact that only spatial frequencies smaller than the optical wavelength will propagate any reasonable distance for far-field detection. As a result, the high spatial frequency content obtained through interaction with objects containing structures much smaller than the wavelength evanescently decays, leading to a long-held notion that imaging such small features requires near field scanning method \cite{Betzig189} to optically image such small object features.

The emergence of super-resolution imaging techniques has shattered the notion that sub wavelength structures cannot be resolved with far-field optical microscopy \cite{Hell1153, Zanacchi2013, SRReview}.  Through the manipulation of excited state populations with nonlinear switching beams, the region of luminescent emission \cite{3DcameraSTED, WE2017} or transient absorption \cite{SATSTED} can be restricted to a region an order of magnitude smaller than a diffraction limited focal spot. These methods require the careful overlap of two laser beams of different colors: an excitation beam and a de-excitation beam.

An alternate method that exploits a simpler experimental strategy only requires a single beam to drive the excited state population into saturation. This method, called saturated excitation (SAX) microscopy \cite{OriginalSAX, Yasunori2018}, is able to produce images that resolve spatial features with a resolution that defeats the diffraction limit. SAX super resolution images are obtained by modulating the total power of the illumination light intensity sinusoidally and measuring harmonics of the input modulation frequency that are recovered from the signal emitted by the object. While SAX microscopy is an elegant laser scanning method that requires only a small modification of a laser scanning microscope, SAX enhancements in resolution improvement suffer from contamination by the shot noise that is present across all harmonics \cite{NonlinearScanningPsaltis}. A further improvement to this method called dSAX \cite{Yasunori2018} extracts the nonlinear signal in a more efficient manner yielding the same resolution enhancements as SAX, but with higher SNR. While this method improves the SNR, it still discards the majority of the energy contained in the signal by separating the higher resolution images according to different orders of non-linear signal extracted.

\begin{figure}[h!]
\centering\includegraphics[width=\linewidth]{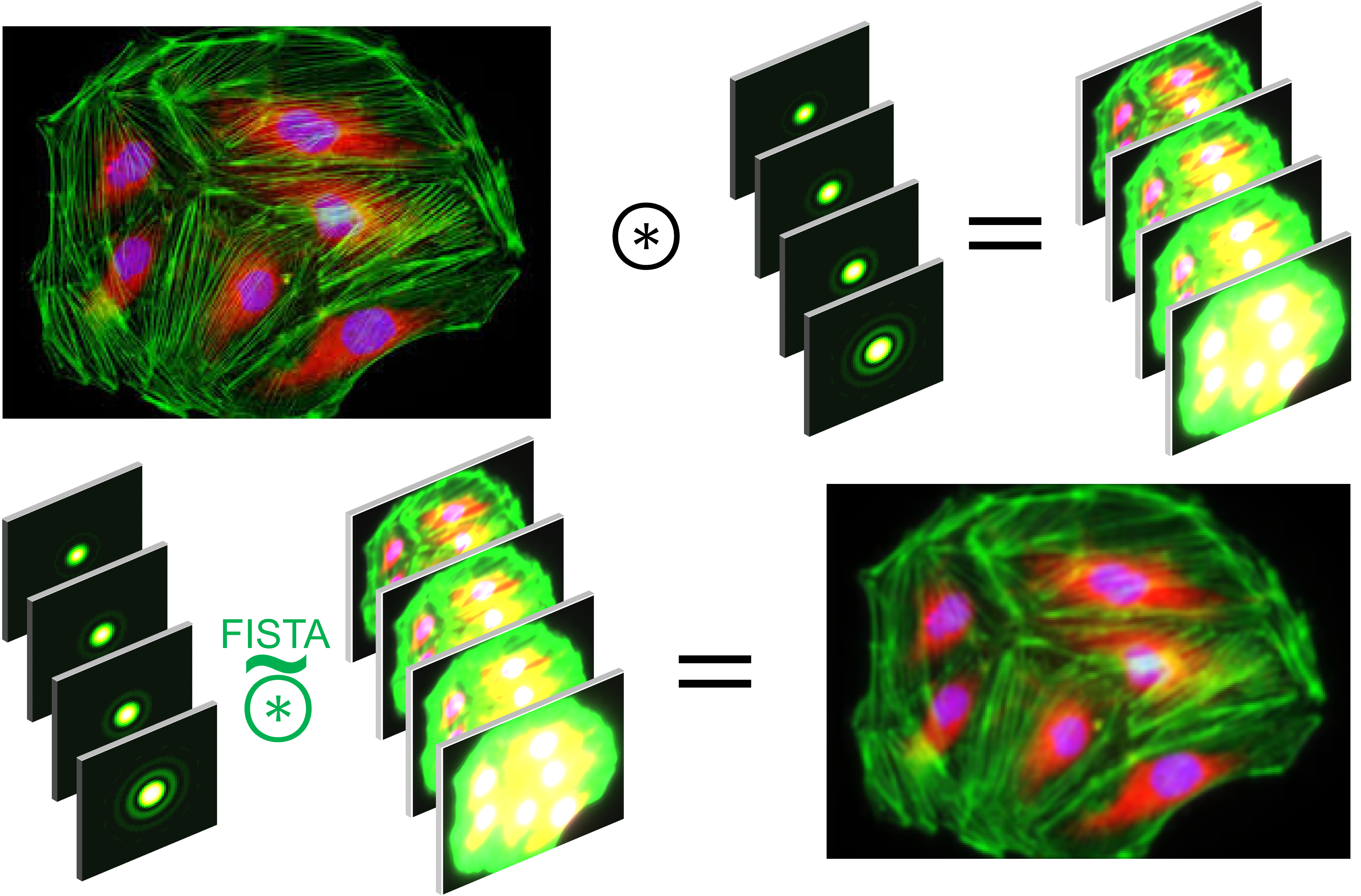}
\caption{Graphical representation of how the data (top) and image (bottom) are formed using CSA (excluding noise). Image courtesy of MicroscopyU \cite{CellImage}}
\label{fig:Concept_fig}
\end{figure}

In this Article, we demonstrate computational saturated absorption (CSA) using a joint deconvolution fusion algorithm called super deconvolution imaging (SDI). \cite{SDIpaper} CSA produces excellent super resolution image quality under conditions where SAX imaging is highly degraded due to corruption by shot noise. In our improved imaging approach, a sequence of laser scanned fluorescent images are acquired under differing levels of saturation of the excited state of the fluorescent molecule. The underlying object is estimated from the set of images that are jointly deconvolved with the set of saturated point spread functions (PSFs) as shown in Fig. \ref{fig:Concept_fig}. The power of the CSA technique is that it does not rely on information only contained at a certain harmonic to get super-resolution information and is able to utilize all the non-linear components together instead of separatly. CSA leverages prior information of the saturation function and iteratively solves for an image which best matches the entire set of data given the model of how the PSF should change with corresponding levels of saturation. This leads to a high resolution, high SNR image without needing to trim out certain portions of the signal that contain the nonlinear response as in SAX and dSAX. At higher levels of saturation, the effective PSF (ePSF) becomes brighter, broader, and steeper, so that the signal-to-noise ratio (SNR) of the data and the steepness of the edges of the ePSF increase. With this strategy, we obtain higher resolution images than allowed by the diffraction limit. Moreover, we show that the image quality obtained by CSA is superior to imaging under similar conditions with SAX microscopy as well as traditional deconvolution techniques.

To appreciate the improvements in super resolution imaging through CSA over SAX, we consider the effect of saturation of a PSF of the two methods. In both cases, we model a laser beam focused to a diffraction limited spot that can be described by the PSF for unaberrated illumination optics, $\PSF_i( \rt ) = J_1( 2 \pi \, \mathrm{NA}_i \, \rt )/( \pi \, \mathrm{NA}_i \, \rt)$. The numerical aperture of the illumination optic is $\mathrm{NA}_i$ and $\rt$ is the radial spatial coordinate $\brh$ that is normalized by the excitation wavelength $\lambda$ and $J_1$ is a jinc function. 
In both CSA and SAX, the illumination beam is used to drive a fluorescent molecule into saturation during excitation. Assuming a 3-level molecular system with a continuous wave (cw) excitation model, the excited state population is given by $e(\rt) = \alpha(\rt)/(1 + \alpha(\rt))$. The local saturation is $\alpha(\rt) = \alpha_0 \, \PSF_i(\rt)$, with the peak saturation value $\alpha_0 = I_0/I_{\rm sat}$ defined as the ratio of the peak illumination intensity, $I_0$, to the saturation intensity of the fluorophore, $I_{\rm sat}$. 

The traditional SAX method for image formation uses information contained in harmonics generated from the nonlinear response of the saturation excitation of the sample. SAX microscopy exploits a sinusoidal temporal modulation of the illumination beam of the form $\alpha(\rt) \, f_m(x)$, where the temporal modulation is of the form $f_m(x) = (1 + \cos(x))/2$, $x = \omega_m \, t$, and $\omega_m$ is the input modulation frequency. Temporal  modulation of the local saturation produces an excited state population that varies in space and time $e_s[\alpha(\rt), x]= \alpha(\rt) \,  f_m(x)/[ 1 + \alpha(\rt) \,  f_m(x)]$. The nonlinear functional mapping imparted by the nonlinear saturated excitation function produces harmonics $q \, \omega_m$, where $q$ is the harmonic order index. Each harmonic separately yields an image, which contains resolution information beyond that of the diffraction limit. With each increasing harmonic, the resolution improves, but the SNR drops dramatically. The imaging model for SAX is generally constructed by considering a Taylor series expansion of the excited state saturation, which generates harmonics of the input modulation frequency $\omega_m$ that lead to an impulse response for harmonics that geometrically scale the the harmonic order, i.e., $\PSF_q(\rt) \propto \PSF_i^q(\rt)$. Unfortunately, the Taylor expansion diverges for relatively small values of peak saturation ($\alpha_0 \sim 0.4$), yet large values of peak saturation are required to produce appreciable SNR in SAX images. The limitation of the Taylor expansion is easily remedied by computing a cosine series expansion amplitude of the excited state population at the q$^{\rm th}$ harmonic of the saturated excited state population $c_q(\alpha) = (2 \, \pi)^{-1} \int_{-\pi}^{\pi} \, e_s(\alpha, x) \, \cos(q \, x) \, dx$. An analytic solution $c_q(\alpha) = (\alpha/2) \, {}_{2}\tilde{F}_{3}\left(1,3/2,2;2-q,2+q; -\alpha \right)$ is expressed in terms of a regularized hypergeometric function ${}_{2}\tilde{F}_{3}$. The PSF for the q$^{\rm th}$ SAX order follows from the expansion coefficient $\PSF_{\rm q}(\rt) = c_q \left[\alpha_0 \, \PSF_i(\rt) \right]$.

 To estimate the SNR, consider a set of $N$ fluorophores with a radiative emission rate of $k_r$ that are localized to a sub-resolution region in space. The maximum signal for the q$^{\rm th}$ SAX order occurs when the peak of the illumination PSF is centered on the fluorescent probe, resulting in a detected photo rate, $\phi_q = \eta_D \, N \, k_r \, c_q(\alpha_0)$, that depends on the level of saturation. SAX signals are subject to multiplexed noise as they are detected in the frequency domain, which means that the shot noise is determined by the average detected photon emission rate that is given by $\phi_0 = \eta_D \, N \, k_r \, c_0(\alpha_0)$, where $c_0(\alpha) = 1 - (1+\alpha)^{(-1/2)}$, indicating that  shot noise rises with increasing saturation. Here $\eta_D$ represents the total detection efficiency, including the detector efficiency, the transmission efficiency through optical components, and the collection efficiency of the objective The peak SNR for the q$^{\rm th}$ order SAX image is $\SNR_q = \kappa_q \, \Upsilon$. Here, $\kappa_q =  c_q(\alpha_0) / \sqrt{c_0(\alpha_0)}$ and $\Upsilon = \sqrt{N \, k_r \, \eta_D \, \Delta t }$, with $\Delta t$ denoting the observation time. For a $\sim 100$ nm diameter sphere with a fluorescent dye and typical numbers of $N \sim 1000$, $k_r \sim 1/3.8$ ns, $\eta_D \sim 0.1$, and $\Delta t = 50 \, \mu$s, then $\Upsilon \sim 1000$. For a peak saturation parameter of $\alpha_0 = 0.4$, then $\kappa_1 = 0.18, \kappa_2 = 0.015, \kappa_3 = 0.0013, \kappa_4 = 1.1 \times 10^{-4}$. Thus, for these numbers, only the first three harmonic orders will rise above the noise threshold even in the shot noise limit. Higher saturation is required to obtain higher harmonics, yet, the PSF for the harmonics will broaden at increased saturation, degrading the improvements in spatial resolution.

In contrast, CSA imaging demonstrated here is able to exploit the high spatial frequency content probed by driving laser scanning microscopy (LSM) into saturation, while simultaneously improving the SNR as higher resolution imaging is scaled with increased peak saturation. In CSA, we record a set of LSM images, each with an increased level of peak saturation of the illumination PSF. A primary advantage of CSA is that the signal at each saturation level is obtained directly from the saturated excitation level, leading to an ePSF given by $\psi(\rt) = (\alpha_0 \, \PSF_i(\rt))/(1 + \alpha_0 \, \PSF_i(\rt))$ that is parameterized by the peak saturation parameter $\alpha_0$. As CSA does not suffer from excess background signal levels, the SNR coefficient for the recorded image at each saturation level is simply determined by the detected fluorescent photons from the peak of the PSF at $\rt=0$, leading to the value $\kappa_{\rm CSA} = \sqrt{\alpha_0/(1 + \alpha_0)}$. A number of ePSFs for a range of $\alpha_0$ values are shown in Fig. \ref{fig:AsymMTFs} d).

CSA jointly exploits the information gained by probing a sample with a set of saturated effective PSFs at several saturation levels. For each image, the incident laser power is adjusted to set the peak saturation $\alpha_{0,s}$ that produces an image with an effective saturated PSF that we denote as $\psi_{s}(\brh)$, and the subscript $s$ labels the image in the set with the corresponding saturation level $\alpha_{0,s}$, which goes from 1 to M. The two dimensional discrete approximation to the ePSF is indicated by the ePSF in bold, $\bpsi_s = \psi_s(x_i,y_j)$ and $x_i$ and $y_j$ are discrete spatial coordinates values. The computational algorithm makes use of a vector form of the discrete ePSF, $\PSFvec_s = [\bpsi_s]_{\fl}$, in which the flattened vector is composed of the columns of $\bpsi_s$ are stacked on top of one another. 

For a fluorescent object with a spatial distribution of fluorophore concentration $c(\brh)$, a recorded image in the set is proportional to $c(\brh) \circledast \psi_{s}(\brh)$ as with any incoherent LSM technique. The discrete approximation to the convolution model can be written as a matrix equation $\vsig_s =\mA_s \, \vc + \vn_s$ that is the discrete two dimensional convolution between $\bpsi_s$ and the discrete object $[\bc]_{i,j} = c(x_i,y_j)$ that has been flattened to a column vector $\vc = [\bc]_{\fl}$. The convolution measurement matrix operator $\mA_s$ is the matrix form of the  discrete convolution integral for an ePSF, $\bpsi_s$. Each row of this matrix represents a shifted version of $\PSFvec_s$. The discrete signal vector, $\vsig_s$, is a two dimensional image which has been flattened into a vector. Noise in the measurement is denoted by $\vn_s$. 

The effective optical transfer function (eOTF) is given by $\Psi_s(k_x,k_y) = \mathscr{F} \{ \psi_{s}(\brh) \}$, where $\kt$ is the conjugate variable to $\brh$ and $\mathscr{F} \{  \}$ is a Fourier transform. The vectorized eOTF, $\OTFvec_s = [\bPsi_s]_{\fl}$, is the flattened form of the discrete eOTF, $\bPsi_s$. As the image formation process for the saturated ePSF is linear and shift invariant, the convolution operator has the form  $\mA_s = \iDFT \, \diag\{ \OTFvec_s \} \, \DFT$, with discrete Fourier transform (DFT) and the DFT adjoint operators denoted by $\DFT$ and $\iDFT$, respectively. As the convolution operator is diagonalized by the DFT, the adjoint of the convolution operator is simply $\mAadj_s = \iDFT \, \diag\{ \OTFvec_s \}^* \, \DFT$, where $^*$ represents complex conjugate. The fact that adjoint operator is equivalent to a cross correlation is exploited to construct a computationally efficient CSA image estimation algorithm. 
The signal spatial frequency vector is thus given by $\vY_s = \OTFvec_s  \circ \vC +\vN_s$, where $\circ$ denotes a element-wise (Hadamard) product between the vectors. Here, $\vC = \DFT \, \vc$, $\vY_s = \DFT \, \vsig$, and $\vN_s  = \DFT \, \vn_s$ are the spatial frequency vectors of the object, signal, and noise, respectively.

\begin{figure}[h!]
\centering\includegraphics[width=\linewidth]{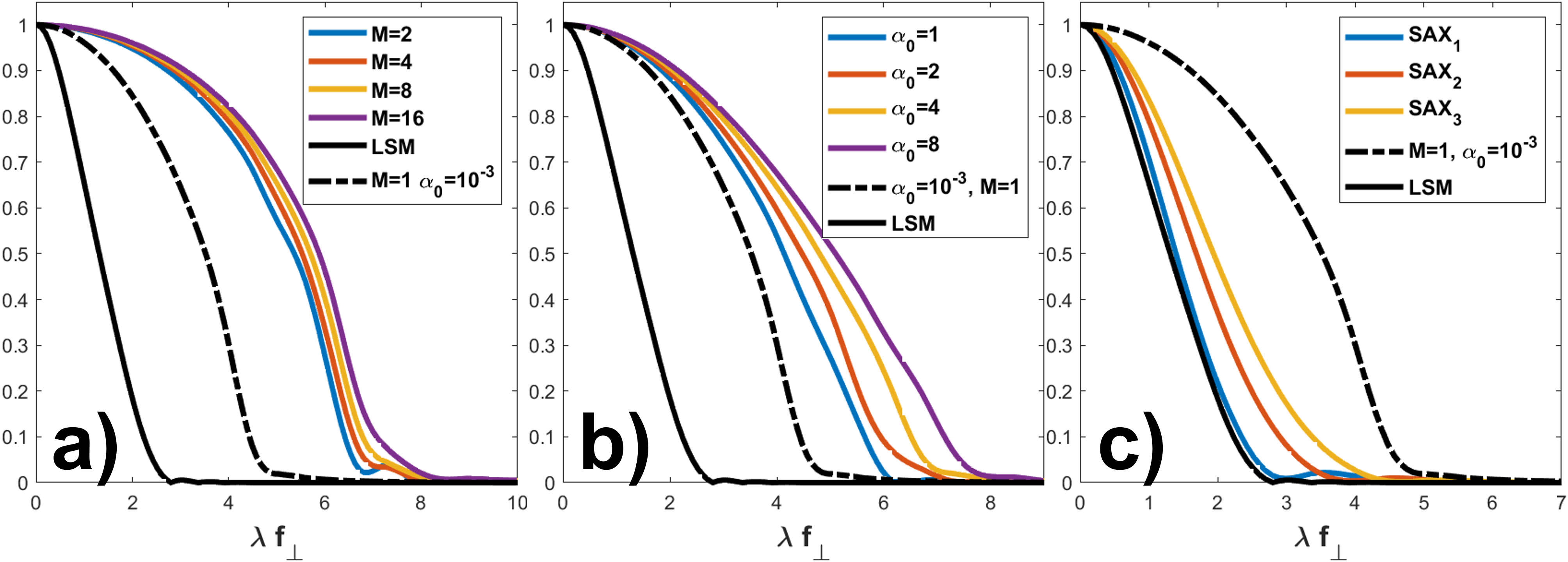}
\caption{These plots show the simulated MTF by direct computation using the pseudoinverse. A point object was reconstructed, which was much narrower than the simulated point spread functions. The Fourier transform of the reconstruction was then taken to produce the MTFs. A computing cluster (ASHA) with large amounts of RAM (192GB) was utilized to carry out these simulations. Plot a) shows the corresponding MTF for increasing the number of point spread functions used in the super deconvolution for the saturation intensity increasing from $.1 \, I_{\rm sat}$ to $4 \, I_{\rm sat}$. The dashed line corresponds to the usual deconvolution using a single PSF with the intensity set to $.001 \, I_{\rm sat}$. The solid black line is the MTF for the laser scanned image without deconvolution. Plot b) shows how the MTF changes with increasing the maximum intensity to different multiples of $I_{\rm sat}$  ($\alpha_0=\frac{I_0}{I_{sat}})$. Each colored curve is the result of the super deconvolution using 15 PSFs rising to the corresponding saturation level. The dashed line is again the deconvolution with a single unsaturated PSF.  Plot c) shows the comparison of MTFs generated with the traditional SAX technique using the same illumination PSF and a saturation level of $\alpha_0=4$. The first, second, and third harmonics are used to generate the MTFs. In all the plots the axis is normalized in terms of wavelength and spatial frequency in order to be unitless. All three simulations used an illumination wavelength of $\lambda=500$ nm with an NA$=1.4$ objective to set the initial resolution.}
\label{fig:MTFs}
\end{figure}

CSA employs a set of data vectors, each taken at a distinct level of peak saturation, $\alpha_{0,s}$, from which the underlying object, $\vc$, is jointly estimated. In the simplest form, we seek to find an optimal object vector $\vc^*$ to the problem formulated as a least mean squared fit
\begin{equation}
    \vc^* := \underset{\vc > 0}{\mathrm{argmin}} \frac{1}{2} \left(\norm{\mA_T \, \vc - \vsig_T}_2 + \norm{\lambda \,  I \, \vc}_2 \right).
    \label{eq:optimzation}
\end{equation}
Here, both the total signal vector, $\vsig_T$, and the convolution operator, $\mA_T$, are concatenations of the full data set of $M$ independent images at separate saturation levels (see Figs. S2, S3), so that for an $N\times$N data scan for each image, each signal vector of length N$^2$ is concatenated to form $\vsig_T$ of length $M \, N^2$. Similarly, $\mA_T$ is a tall and skinny matrix with dimensions of $MN^2 \times N^2$ (each $\mA_s$ is stacked on top of one another vertically, as shown in Fig. S2). To illustrate the CSA principle and to estimate the PSF and OTF of the result of super deconvolution, we compute Eq. (\ref{eq:optimzation}), with a very small $\lambda$ value (this avoids large values in the reconstruction due to small eigenvalues of $\mAadj_T \, \mA_T$). We directly obtain the LMS solution by by computing the regularized Moore-Penrose pseudoinverse of the concatenated data and convolution operators, which yields the solution $\vc^* = (\mAadj_T \, \mA_T+\lambda \, I)^{-1} \, \, \mAadj_T \vsig_T$, where $^{-1}$ indicates the matrix inverse. The results of noise-free simulations of a sub-resolution  point object to estimate the PSF and OTF are presented in Fig. (\ref{fig:MTFs}). Even in a noise-free case, CSA outperforms both a simple linear devolution and conventional SAX imaging by providing much better spatial frequency support for image formation. Asymptotic behavior of the OTFs with very high levels of saturation are further explored in Fig. \ref{fig:AsymMTFs}. 

Computation of the pseudoinverse provides a direct solution to the super deconvolution problem for CSA, but is infeasible for realistic image sizes. For an image size of 256x256 and a set of 15 images (each at a different saturation level) with zero padding, the matrix occupies $\sim$ 512GB of data, which exceeds the available memory in most computers. This computational hurdle can be significantly reduced by carrying out the operation of $\mA_T$ and $\mA_T^\dagger$ equivalently in terms of Fourier transforms -- eliminating the need to store a large concatenated Toeplitz matrix, $\mA_T$, in memory. Due to computational constraints, we solve the CSA problem with an iterative optimization algorithm called FISTA (fast iterative shrinkage thresholding algorithm). FISTA is a regularized form of a gradient descent optimization algorithm, and as such, we need a current guess and the gradient of equation (\ref{eq:optimzation}). The key to handling large data sets with such an algorithm is to find a method of computing the gradient value for each iteration without requiring instantiating $\mA_T$ \cite{Antipa:18}. 

For efficient computation, the gradient term in the FISTA algorithm is computed without inverting and instantiating the large matrices that arise in the CSA problem. To do so, the gradient, $\vgr$, of the Eq. (\ref{eq:optimzation}), which for a single image deconvolution is given by $\vgr_s = \mAadj_s \, \left(\mA_s \, \left[ \vc - \vsig_s \right] \right) + \lambda^2 \, \vc$ is computed through FFT operations with $\vgr_s = \iDFT \, \left( \OTFvec_s^* \circ \DFT \, [\iDFT ( \OTFvec_s \circ \vC)] - \vsig_s \right)$. As shown in the supplements, the efficient gradient computation can be extended to the concatenated set of saturated measurements by looking closely at the operation of the joint convolution operator $\mA_T$ and its adjoint $\mA_T^\dagger$ and comparing to the calculation of the convolution and cross correlation using Fourier transforms. The convolution of the concatenated set of progressively saturated ePSFs, $\bpsi_T$, and the concentration map of our image $\bc$ can be calculated $\bpsi_T \circledast \bc = \iDFT \, \left( \DFT \, \bpsi_T  \circ \DFT \, \bpad[\bc] \right)$. The $\bpad()$ operation zeropads the object $\bc$ such that it becomes the same size as $\psi_T$ ($\psi_T$ is $MN \times N$ and $\bc$ is $N \times N$). The padding operation allows the Hadamard product to be carried out and gives the same result as $\mA_T \, \vc$ after being flattened. The adjoint operation $\mA_T^\dagger$ operates on an array which is the same size as our data vector $\vsig_T$ and outputs a vector the size of $\vc$.

The adjoint of a convolution is a cross correlation which can be calculated using Fourier transforms as $\psi_T \star \bv_T= \iDFT \, \left( \DFT \,  \bpsi_T^* \circ \DFT \, \bv_T  \right)$ ($\star$ represents the cross correlation and $\bv$ is a dummy array the same size as $\psi_T$). The problem here is that the output of this calculation is not the same size as $\mA_T^\dagger \, \bv_T$. This discrepancy is due to the limited number of columns of $\mA_T$. The operation of $\mA_T^\dagger$ only has enough rows to shift $\psi_T$ with respect to $\bv_T$ one image width when carrying out the cross correlation. As shown in the supplements, this means the $\mA_T^\dagger$ operation only returns the central part of the cross correlation of the concatenated arrays. This is equivalent to $\sum_{s=1}^M \psi_s \star \bv_s$. The full cross correlation returns the concatenation of the sum of different combinations of images in the set given as  $  \iDFT \, \left( \DFT \,  \bpsi_T^* \circ \DFT \, \bv_T  \right) =\left[\psi_1 \star \bv_1 \biggr\rvert \sum_{j=1}^2 \psi_j \star \bv_j \biggr\rvert \sum_{j=1}^3 \psi_j \star \bv_j \biggr\rvert \sum_{j=2}^3 \psi_j \star \bv_j \biggr\rvert \psi_3 \star \bv_3 \right]$ (for M=3). The required adjoint is contained only in the central image of this full cross correlation. 

\begin{figure*}[h!]
\centering\includegraphics[width=\linewidth]{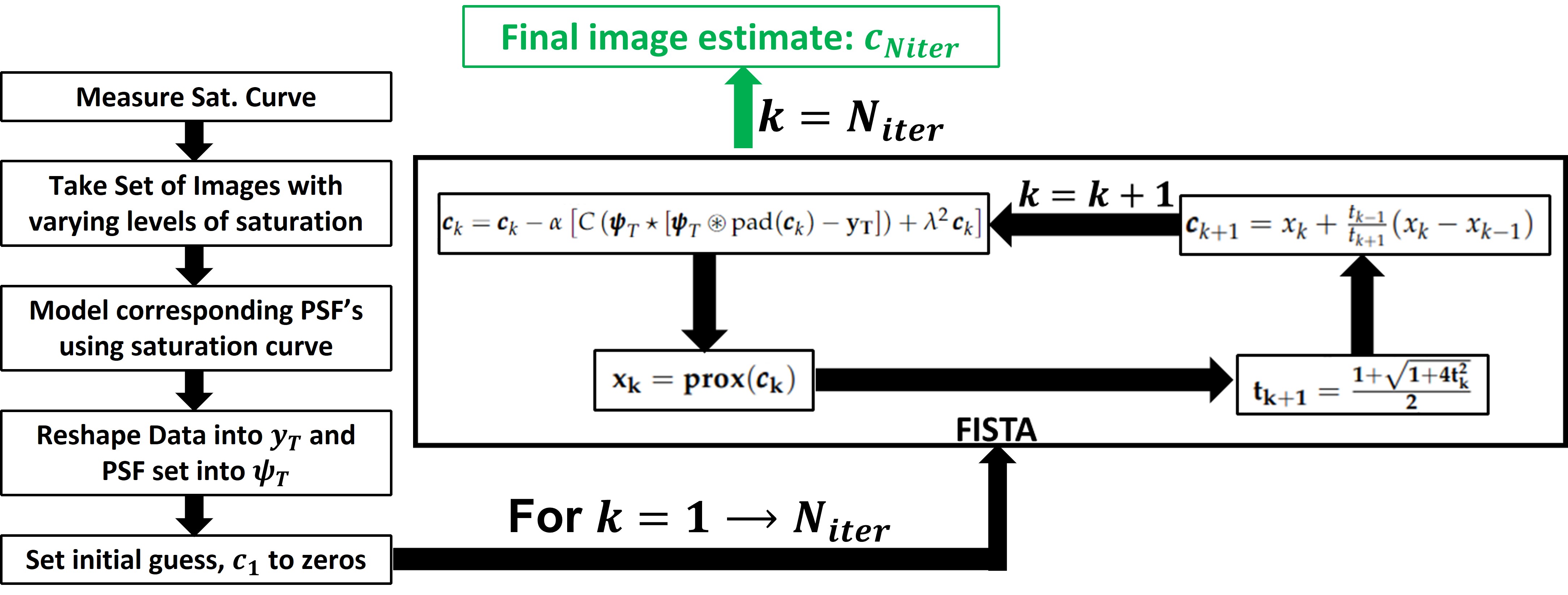}
\caption{Block diagram showing each step of CSA imaging as well as the FISTA algorithm.}
\label{fig:BlockDiagram}
\end{figure*}

We now can calculate the operation of $\mA_T$ and $\mA_T^\dagger$ using efficient FFTs without holding $\mA_T$ in memory using  $\mA_T \, \vc = \left[ \iDFT \, \left( \bPsi_T \circ \DFT \, \bpad[\bc]  \right) \right]_{\fl}$ and $\mA_T^\dagger \,  \vv_T= \left[ C \left( \bPsi_T^* \circ \DFT \, \bv_T \right)\right]_{\fl}$. Here $C(\cdot)$ is a cropping operator that denotes taking the central part of the array and $[\cdot]_{\fl}$ describes taking the 2D array and flattening it into a vector. Using this description, we can now carry out the super deconvolution of the set of data and saturated point spread functions using FISTA while avoiding storing large matrices in memory. Typical run times to perform the joint deconvolution range from 5-15 minutes depending on the size of data and required number of iterations. This produces a single image that combines information from each LSM image to synthesize a high SNR super resolution image. For a comprehensive overview and detailed explanation of the CSA algorithm see the supplemental information that accompanies this paper. Pseudocode for the algorithm is shown in algorithm S2 in the supplements and a block diagram is shown in Fig. \ref{fig:BlockDiagram}.

\begin{figure}[h!]
\centering\includegraphics[width=\linewidth]{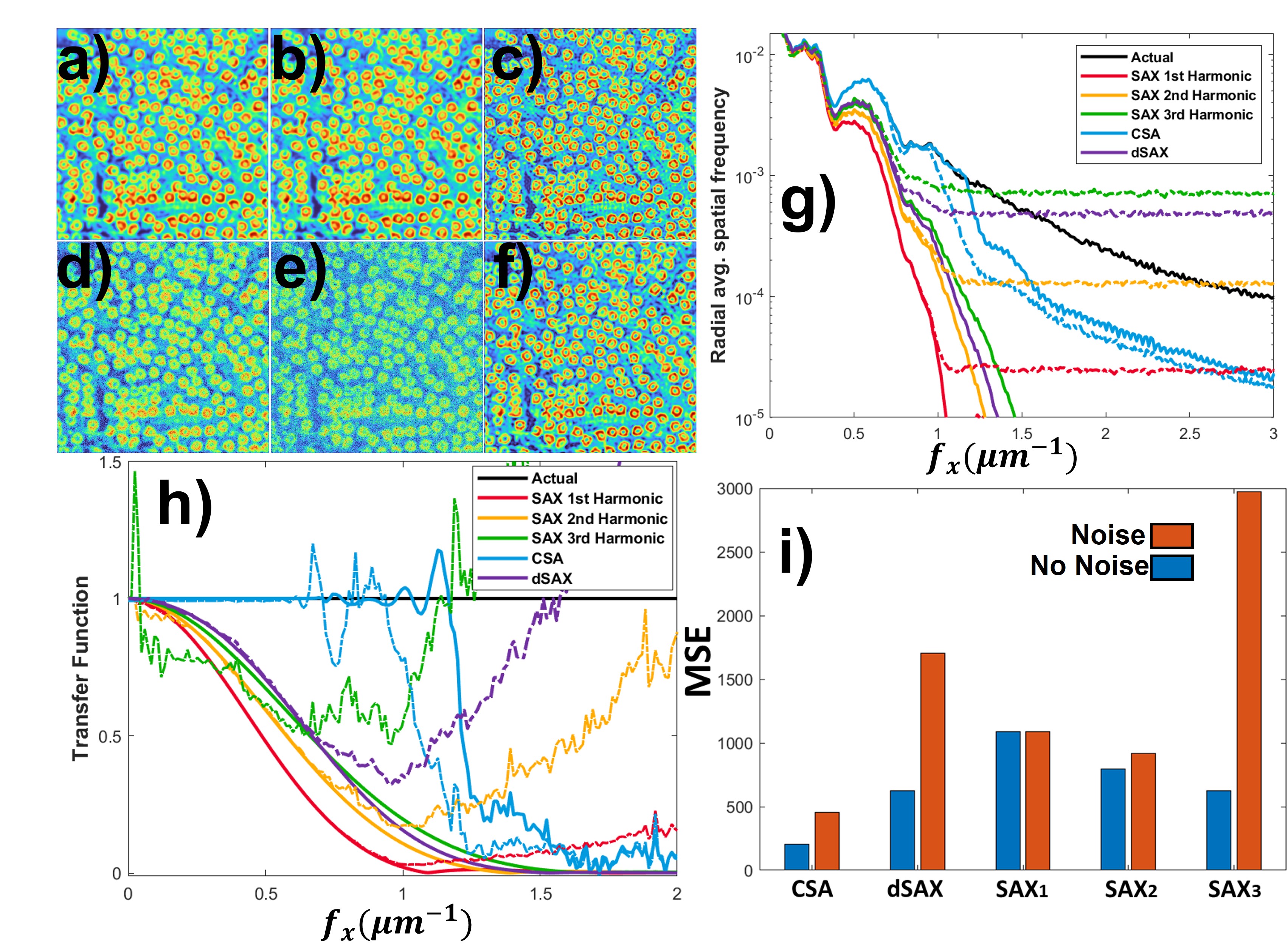}
\caption{Simulations comparing the imaging performance of the dSAX (a, d), SAX (b, e) and CSA (c, f) imaging techniques with and without noise. Images a),b), and c) show the reconstructions without noise and d), e), f) with additive Gaussian noise and Poisson noise present in the simulated data (Mean photon count of 25000 and additive Gaussian noise with standard deviation of 1\% of the maximum value). Images a) and d) show the reconstructed images using dSAX extracting the third order non-linear signal \cite{Yasunori2018}. Image b) and e) show the reconstructed images using SAX demodulated at the third harmonic. Images c) and f) use the CSA technique with a set of 15 images with the intensity going from .01 to 4$I_{\rm sat}$. The data is simulated with a nuclear pore complex image \cite{NuclearPoreImage}. Plot g) shows the radial average of the spatial frequency content of the reconstructed images with each normalized to the DC value. In plots g) and h) the solid lines indicate the reconstructions without noise present and the dashed lines represent reconstructions with noise present. Plot h) shows the transfer function of the reconstructed images. This is found by dividing the Fourier transform of the reconstructed image by the Fourier transform of the true image and then taking the radial average. Bar graph i) shows the mean squared error of each of the reconstructions (MSE). SNR values were also calculated for CSA (23.3dB), dSAX (18.1dB), and SAX demodulated at first through third harmonics (38.17dB, 21.3dB, 8.7dB)}
\label{fig:CellRecons}
\end{figure}

To benchmark the expected performance of the CSA algorithm, we performed simulations, the results of which are shown in Figure \ref{fig:CellRecons}. As discussed earlier, CSA is able to fully exploit the improved high spatial frequency object information because this strategy does not suffer from multiplexed shot noise that buries the higher harmonics in traditional SAX imaging. The top left set of images in  Figure \ref{fig:CellRecons} displays the results of our simulations, where the left column shows dSAX with and without noise (d and a respectively), the middle column of images are traditional third order ($q=3$) SAX images, in the noise-free b) and  with noise in e).  The right column of images are those for CSA, with c) showing the noise-free result and f) showing the result when noise is present. While the bottom row of these images are degraded compared to the top row, CSA super resolution imaging is significantly more robust to noise, and produces a much higher quality image. Shown in Fig. \ref{fig:MTFs}a) the number of point spread functions used does not show a dramatic difference in OTF support. In the simulations fifteen point spread functions were used, but similar results should be expected using a lower amount. While the superior image quality is evident in Fig. \ref{fig:CellRecons}(f), the image itself provides no quantitative argument for the superior image quality. 

To evaluate the spatial frequency information content of the reconstructed images, the radially averaged spatial frequency content of the images are shown in Fig. \ref{fig:CellRecons}(g). The true object radial spatial frequency distribution is denoted by the solid black line. Noise free reconstructions for a maximum peak saturation parameter of $\alpha_0 = 4$ are shown in solid lines for three harmonics of SAX, dSAX and for CSA.  Image reconstructions in the presence of both additive Gaussian (standard deviation of 1\% of the maximum value) and Poisson noise (mean photon count = 25000) are indicated by dashed lines of the same color. CSA shows robust behavior in the presence of noise, whereas the limitations of the multiplexed shot noise is evident in the constant values of spatial frequency amplitude at high spatial frequencies that is determined by the value of $c_0$ in the cosine expansion. In the case of SAX, we see that the cutoff spatial frequency for imaging depends on the noise level. As the noise level rapidly rises with higher SAX orders, the cutoff spatial frequency is reduced. These particular values of cutoff depend on the noise level in the measurement, which in turn follows the average value of the fluorescent emission. Thus, traditional SAX imaging is likely limited to bright objects. The dSAX method was also compared which extracts the third order non-linear signal from the same saturated images used for CSA. The result is that with no noise present both dSAX and the third harmonic from SAX yield nearly identical results. When noise is introduced the dSAX method yields much higher SNR. Even though dSAX gives much better SNR than SAX it still relies on the extraction of the third order non-linear signal which suffers from the same multiplexed noise. While CSA is not subject to multiplexed noise, we see that the amplitude of the spatial frequency content is attenuated at high spatial frequencies when noise is present. 



The quality of the estimated images can be quantified by computing the signal-to-noise ratio (SNR) with the formula $\mathrm{SNR} = 10 \log_{10}  \left[ \sum\limits_{i=1}^N \hat X_i /  \sum\limits_{i=1}^N (\hat X_i-X_i) \right]$ and the mean squared error (MSE) with the formula $\mathrm{MSE} = N^{-1} \, \sum\limits_{i=1}^{N}(X_i-\hat X_i)^2$, where $N$ is the total number of elements, $X$ is the true image and $\hat X$ is the observed image. These formulae benefit from the fact that for the simulations the true object is known. Here we see that the first-order SAX image, which is similar to a conventional LSM image, is robust to this particular level of noise. However, this first order image has low spatial resolution compared to CSA and higher SAX orders. The second order SAX image is mildly affected by the noise, whereas the third order SAX image is severely degraded. The CSA image SNR and MSE are mildly degraded by the noise, however, the SNR is significantly higher and the MSE is significantly lower for the CSA image compared to all SAX image orders. Compared to dSAX CSA also results in higher SNR and spatial frequency content as observed in both the radial average of the spectrum and transfer function compared to the ground truth. While further study is required to fully explore the impact of noise, these conditions clearly show significant benefits for CSA imaging. 

\begin{figure}[h!]
\centering\includegraphics[width=\linewidth]{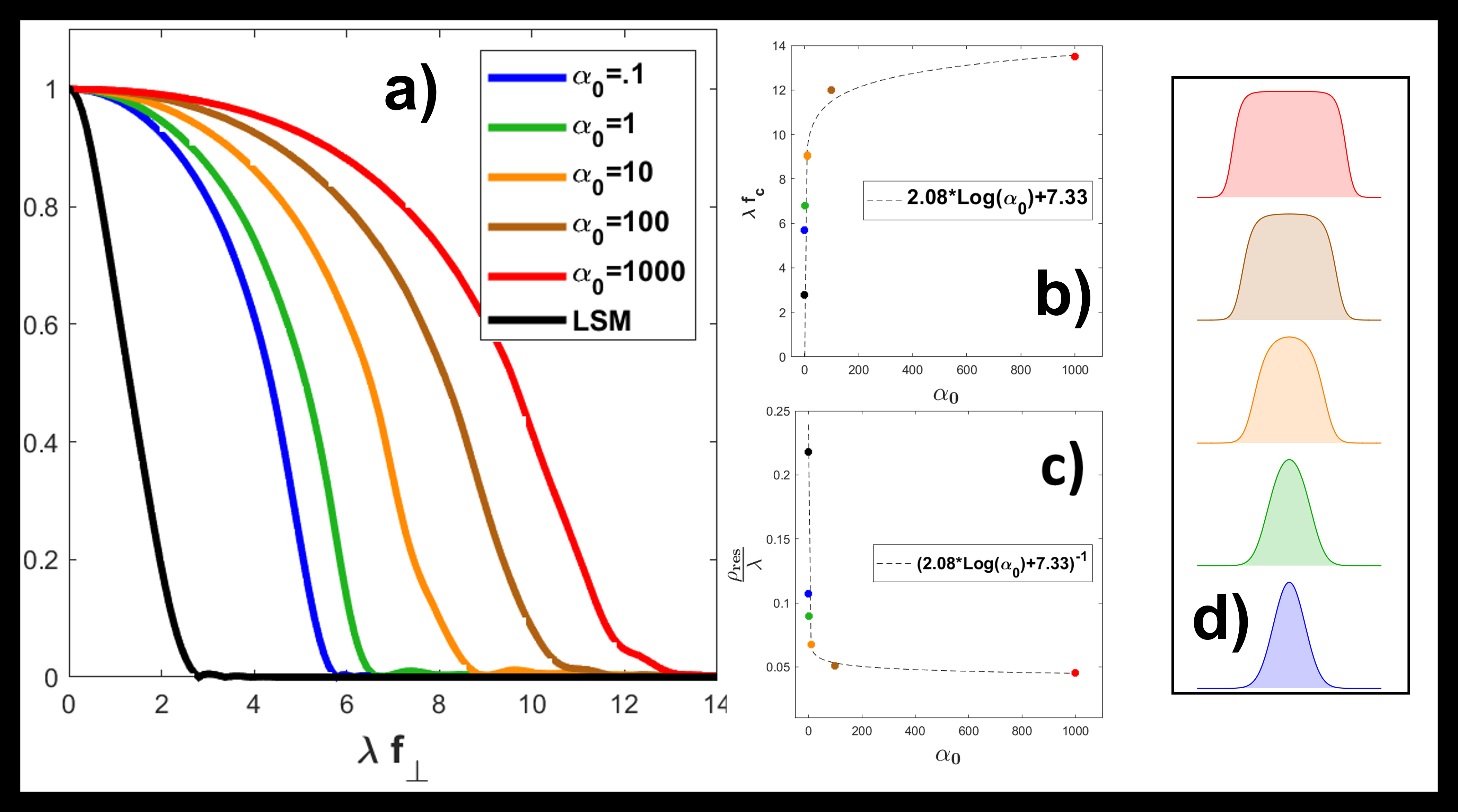}
\caption{Plot a) shows the expected MTF for increasing levels of saturation in the asymptotic limit. Plots b) and c) show how the expected resolution trends with max saturation level. Plots a), b), and c) are normalized in terms of wavelength so they are all unitless. Panel d) shows how the PSF changes shape according to the maximum saturation level of each simulation. Each MTF is generated from using the CSA technique with a set of fifteen PSFs evenly sampling the saturation curve starting from $.01 \, I_{sat}$ to the maximum saturation level. The plots to the right show the PSFs at increasing levels of peak saturation ($\alpha_0 = I_0 /I_{sat}=[.1, 1, 10, 100, 1000]$ respectively)}
\label{fig:AsymMTFs}
\end{figure}

\begin{figure*}[h!]
\centering\includegraphics[width=\linewidth]{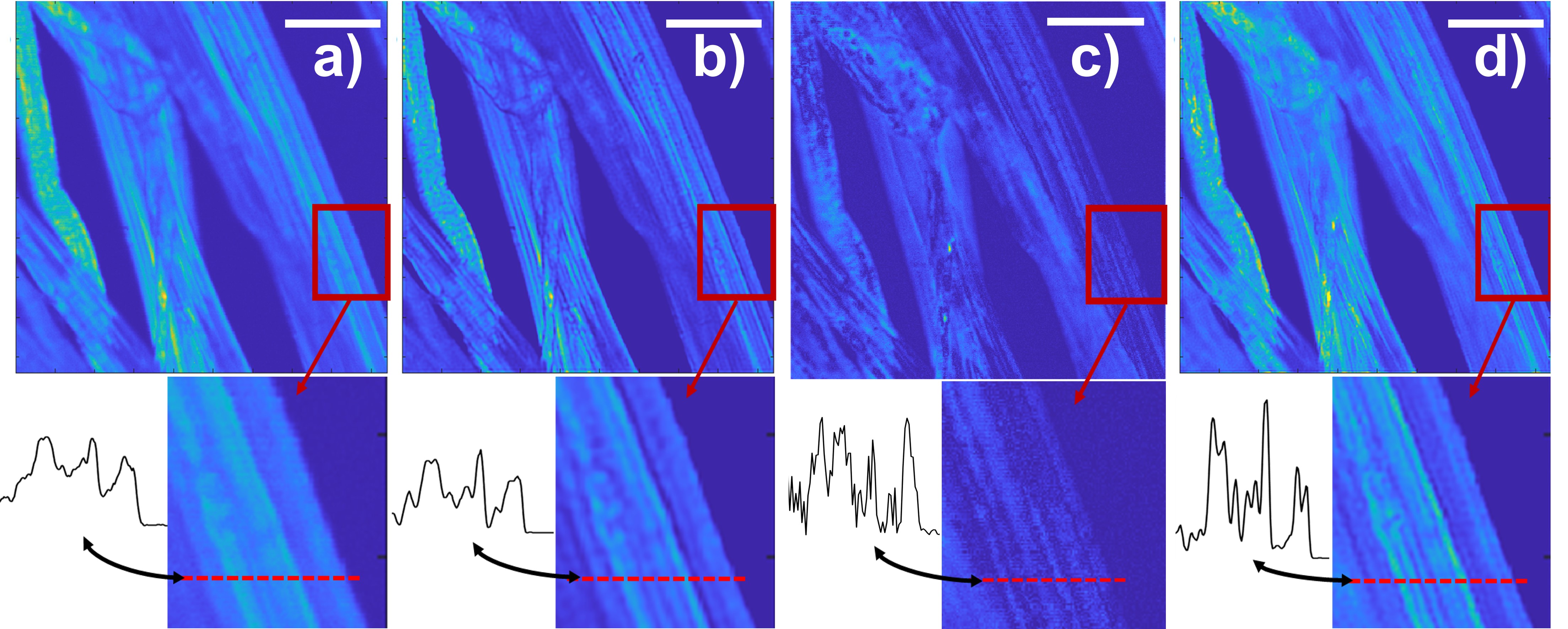}
\caption{Image (a) shows a laser scanned image with an input intensity of $0.29 \alpha_0$. Image (b) is the resultant deconvolution of the laser scanned image to the left using FISTA. Image (c) shows the resultant image using dSAX \cite{Yasunori2018}. Image (d) is the resultant image using CSA with all four laser scanned images using FISTA. Bellow each image shows a zoomed in portion of each corresponding reconstruction along with a line out. This clearly shows a resolution enhancement in the case of CSA since more fibers are able to be resolved. This is especially evident near the right most edge of the fibers where only the CSA reconstruction is able to distinguish that there are two sub-resolution fibers which run parallel to each other along this edge. The dSAX image shows improvement in resolution similar to the deconvolution, but the SNR is poor in comparison to the CSA image. Scale bar is 25 $\mu m$.}
\label{fig:Reconstructions}
\end{figure*}

Experimental CSA images are shown in Fig. (\ref{fig:Reconstructions}). The experiments are conducted with $\sim 250$ fs pulses centered at 1035 nm (Y-Fi NOPA, ThorLabs Inc.). The peak saturation parameter, $\alpha_{0,s}$, is controlled by varying the average power of the excitation beam with a constant amplitude RF driving signal applied to an acousto-optic modulation (AOM). The illumination beam is directed into a laser scanning nonlinear microscope \cite{Young:15}. For this experiment two-photon absorption of fluorescein dyed fibers was used. There is no change in the algorithm from linear absorption to two photon absorption for CSA and dSAX techniques. The only modification for CSA is to ensure the model for the saturated point spread function is adapted to account for a slight change in the saturation function. This is described in detail in section 6 of the supplemental information. The emitted two-photon fluorescence from the sample is collected in the forward direction with a photo-multiplier tube (PMT) after being passed through a dichroic filter to reject the pump light. Saturation curves are measured by recording the PMT signal as a function of incident illumination power that is rapidly varied applying a ramp function to the RF modulation signal amplitude of the AOM -- allowing accurate estimation of the point spread function shape as intensity values reach different saturation levels. 

The fluorophore is excited through two photon absorption with the pulsed laser source. This excitation produces a saturation curve that is shown in Fig. S7 of the supplements. Accurate modeling of each saturated PSF requires that the saturation curve be well characterized. Measured fluorescent saturation data are fit to the pulsed two photon excitation model given in Eq. S16 of the supplements. The fit to the experimental data is used as the nonlinear function map that transforms the measured low intensity linear PSF to the ePSF at a given experimental saturation level to estimate the saturated ePSFs used in the super deconvolution process. The effective linear input PSF is estimated from an image of a 100-nm diameter fluroescent nanodiamond under conditions of weak, i.e., unsaturated, excitation. Saturated ePSFs for the super deconvolution model are synthesized from the saturation curve and the illumination PSF.

Images of the sample, composed of tissue fibers stained with fluorescein dye, were taken at four different power levels reaching a maximum peak saturation level of 3 times the two-photon saturation intensity, i.e., $\alpha_0^{(2)} =3$. These images are used to obtain a saturation curve by rapidly measuring the fluorescent signal with the excitation beam parked at one position in the sample. The measured fluorescent signal data points are fit to Eq. S16, that expresses the the mean fluorescence signal, $F$, for two photon emission assuming the pulse duration is much shorter than the fluorescence lifetime and time between pulses, $I_{\rm sat}^{(2)}$ is the two photon saturation intensity. With the saturation curve based on a fit to the experimental fluorescent power saturation curve, the CSA algorithm can be applied to the image data. The images are first aligned with a cross correlation to remove any  spatial drift accumulated between the scans. The image alignment is performed by minimizing the cross correlation amplitude between a selected reference image and the remaining images in the stack. Then, the a set of saturated ePSFs are computed using the experimentally derived linear PSF measurement and the two-photon saturation curve function. Both the images and the ePSFs are normalized to unity at their peak values and run through the CSA deconvolution algorithm, which a FISTA optimization routine that is used to estimate a single high resolution image. The experimental reconstructed images are shown in Fig. (\ref{fig:Reconstructions}). For comparison, Fig. (\ref{fig:Reconstructions})(a) shows the LSM image for a relatively weak excitation at $\alpha_0^{(2)} = 0.29$. The simple deconvolution of this image shown in Fig. (\ref{fig:Reconstructions})(b) shows that higher resolution features can be extracted from the two-photon fluorescence image of fluorescein dyed fibers. A reconstructed image solving for the third order non-linear signal using the dSAX technique is shown in (\ref{fig:Reconstructions})(c. A CSA image obtained from a set of four saturation levels with $\alpha_0^{(2)} = {{.29, .8807, 1.81, 2.98}}$ produces a higher resolution image with better SNR. For comparison, a zoomed in portion of each image is shown with plots of a line-out of a section of the image. This clearly shows a resolution enhancement afforded by the CSA technique shown in Fig. (\ref{fig:Reconstructions})(d) with the CSA image resolving more fibers than the LSM, deconvolved or dSAX images. The LSM image shows a relatively high shot noise level. The dSAX image shows a nice improvement in resolution, but the SNR is quite low. Single linear deconvolution helps suppress this noise and brings out some of the high spatial frequency information in the image. CSA brings out more information than the linear convolution and dSAX across the spatial frequency band since it utilizes information from many orders of the non-linear signal simultaneously, which highlights the improved imaging performance using CSA.

To summarize, we have introduced a new super resolution optical imaging modality in which we exploit information from a set of images, each acquired with a distinct ePSF. In this work, the set of ePSFs, each of which corresponds to a fluorescent image acquired at increasing levels of saturation of the fluorescent excitation, are jointly deconvolved to produce a super resolution image. While the excitation of the fluorescent excited state is nonlinear in the case of saturation, we may define an effective PSF that follows a linear image formation model that is determined by the saturation curve of the excited fluorescent molecule. As the saturation level increases, these ePSFs become steeper, wider, and brighter. As a result, the eOTF exhibit both higher SNR and higher spatial frequency support at higher saturation levels. We have developed a computationally efficient strategy to jointly solve a super deconvolution problem by combining the all of the information represented in the spatial frequency diversity across the set of measured images with the set of ePSFs. This computational saturated absorption (CSA) strategy combines all of the information acquired from the set of measurements. Each measurement spans a range of transverse spatial frequencies. The weakly saturated images have a narrower range of spatial frequency support, and thus exhibit higher SNR in the image data at low transverse spatial frequencies. By contrast, the higher saturation level images have higher signal overall, but that signal is spread across a much broader spatial frequency range. The super deconvolution requires that the estimated image simultaneously satisfy all of the measured image data. As a result, the high SNR at low spatial frequencies provided by images with low saturation helps to stabilize the higher spatial frequency content obtained with the saturated images. Consequently, CSA yields an improved spatial resolution as well as higher SNR images than is possible with conventional LSM, SAX, dSAX or even from deconvolution of either LSM or SAX images. Moreover, Fig. \ref{fig:AsymMTFs} shows that the resolution improvements continue to scale as the peak saturation increases. Obviously the improvements can not scale indefinitely, but some fluorescent and luminescent systems can reach extremely high saturation levels while remaining well below any damage threshold. Indeed, even in the case of organic dyes that can tolerate GW/cm$^2$ peak intensity levels, $\alpha_0 \sim 3000$ is possible. In the case of systems with similar absorption cross sections, but long lifetimes, such as rare earth ions and photoswitchable proteins, the peak saturation level can exceed $10^5$, suggesting extremely fine spatial resolution imaging is possible. Further limitations of this technique are its reliance on relatively high saturation levels, which can cause photobleaching or damage in some samples. Another limitation is the requirement of precise knowledge of the point spread function of the system and the samples saturation function.   

\begin{acknowledgments}
We acknowledge funding support from the Chan Zuckerberg Initiative, the National Institute of Health (NIH) (R21EB025389, R21MH117786) and the Department of Energy (DE-SC0017200). J. Squier is supported by the National Science Foundation (NSF)(1707287).
\end{acknowledgments}

\appendix

\section{Supplemental Information}

These supplements present the derivation of the computational saturated absorption (CSA) microscopy super devolution algorithm which is used to achieve robust super-resolution laser scanning microscopy using a set of point spread functions with diversity in saturation of the material excitation.

Optical imaging systems inevitably capture coarser spatial features than may be present in a particular specimen. The spatial resolution is limited by the properties of the illumination beam, such as wavelength, spatial frequency bandwidth as established by illumination and collection optics, aberrations, and optical coherence. The combined effect of the properties of the illumination light and the optical microscope system produce, within an aplanatic limit, a system model that may be described with as linear and shift invariant \cite{mertz_2019}. As our paper is focused on fluorescent imaging, we restrict our discussion to an incoherent image transfer model where the spatial impulse response of the imaging process is given by a point spread function (PSF). We note that while our specific experimental implementation uses two photon absorption to excite fluorescent molecules in the specimen, our approach generalizes to any system that can be described with a set of diverse PSFs. As is made evident in the paper, the PSFs produce a blurry image that degrades the potential spatial resolution. 

As the forward model for the image is represented by a convolution, knowledge of the PSF, or in our case, a set of PSFs can be exploited to undo the blurring of the image. Such a deblurring process is called deconvolution \cite{DeconvolutionIEEESigProcessing, SAXdeconv, doi:10.1063/1.5039567}. Image deconvolution is widely used in many fields and rose to prominence after the launch of the Hubble telescope that produced low-quality blurred images as a result of faulty optics in the telescope \cite{Hubble}. Deconvolution algorithms exploit the simple mathematical structure that the forward spatial convolution is a product of the desired spatial frequency distribution of the specimen spatial variation and the spatial Fourier transform of the PSF, which is known as the optical transfer function (OTF). Such a description is cripplingly na\"ive because noise in the measurement destabilizes such a simple inverse solution \cite{ThiebautDeconvolutionInAstronomy, Sibarita2005}. As such, image deconvolution employs methods of inverse problems, which often leads to a need to seek iterative solutions through an optimization algorithm \cite{Antipa:18}. We use such a method in our paper, and in these supplements, we provide a tutorial-level description of our iterative algorithm. 

Image deconvolution methods invariably use a form of regularization for solving the inverse problem \cite{ThiebautDeconvolutionInAstronomy, Sibarita2005}. Regularization produces spatial blurring due to a reduction in spatial frequency content that contributes to the final estimated image. In the absence of noise, the best spatial frequency support (a robust way to gauge the resolving capability of an imaging system) will extend up to the highest spatial frequency that is passed from the specimen to the image by the experimental system. Deconvolution is able to boost the amplitude of the spatial frequency transfer from specimen to image, but this capability is strictly limited by the presence of noise. Data with low signal-to-noise ratios (SNR) often suffers from reduced spatial frequency support \cite{Arigovindan17344}. The use of image priors (i.e., prior information) can be used to extend extracted spatial frequency information beyond the spatial frequency support of the imaging system, but with the caveat that a mismatch between the prior information and the data arising from the specimen will inject errors and artifacts into the estimated image \cite{BERTERO2003265}. The most useful version of super resolution microscopy that relies on image priors is localization microscopy \cite{LMreview, DeconvSR}.

\section{Notation}

The notation for discussing the CSA super deconvolution algorithm involves multiple representations of similar objects. In each case, we use a consistent notation. When speaking of continuous function, such as the object, $c$, we will use conventional script, whereas the two dimensional discrete matrix representation of those functions will be represented by a bold variable, $\bc$, and finally the flattened version of the matrix representation will be written as a column vector in the form $\vc$. Specifically, $c$ spans the real continuous domain, whereas for a discretized object on an $N \times N$ grid, the flattened vector $\vc$ will be a column vector of length $N^2$. This notation is extended to all quantities discussed here and the notation is summarized in the tables below.
\begin{tabular}{ |p{3cm}||p{2.1cm}|p{3cm}||p{2.1cm}|  }
\hline
 \multicolumn{4}{|c|}{\bf CSA super deconvolution operators} \\
 \hline
 Convolution & $[\cdot]\circledast [\cdot]$ & Cross correlation & $[\cdot]\star[\cdot]$ \\
 Hadamard product & $[\cdot]\circ[\cdot]$ & Complex conjugate & $[\cdot]^*$ \\
 Adjoint & $[\cdot]^\dagger$ & Inverse & $[\cdot]^{-1}$ \\
 Fourier transform & $\mathscr{F}\{\cdot\}$ & Inverse Fourier transform & $\mathscr{F}^{-1}\{\cdot\}$ \\
 DFT & $\DFT\{\cdot\}$ & iDFT & $\DFT^{-1}\{\cdot\}$ \\
 DFT matrix& $\DFT$ & iDFT matrix& $\iDFT$ \\
 Zero padding & $\bpad [\cdot]$ & Crop & $C\ [\cdot]$ \\
 Flatten & $[\cdot]_{\fl}$ & Diagonal matrix (put vector on diagonal)& $ \diag\{\vec{[\cdot]}\}$ \\
 Convolution matrix & $\mA_s$ & Super convolution matrix & $\mA_T$ \\
 Concatenation & $\bigg[\ [\cdot] \ \bigg| \ [\cdot]\ \bigg]$ & $L_2$ norm & $\norm{\ [\cdot]\ }_2$ \\
 \hline
\end{tabular}

The various operators used in this work are also listed in these tables. For example, the convolution matrix operators $\mA_s$ and  $\mA_T$ are size $N^2\times N^2$ and $MN^2\times N^2$ respectively, but can generally be larger when zero padding is added which is best practice. The notation denoting the DFT and the DFT matrix and their inverses have subtle differences. The discrete Fourier transform (DFT) is denoted $\DFT\{\cdot\}$ where the DFT matrix operator is $\DFT$. These both carry out the same calculation, just on differently shaped data. The DFT matrix $\DFT$ operates on a vector which is a flattened two dimensional array and outputs another flattened vector. The DFT $\DFT\{\cdot\}$ operates on a two dimensional array and outputs a two dimensional array which is its two dimensional discrete Fourier transform.

\begin{tabular}{ |p{3cm}||p{1.5cm}|p{3cm}|p{2.25cm}|  }
 \hline
 \multicolumn{4}{|c|}{\bf CSA super deconvolution variables} \\
 \hline
 Variable name & Notation & Construction & Domain or size \\
 \hline
 \multicolumn{4}{|c|}{Object} \\
 \hline
 Concentration   & $c(x,y)$    &  &  Real space \\
 Discrete object array &  $\bc$  & $[\bc]_{i,j} = c(x_i,y_j)$   & $N \times N$ array\\
 Flattened object & $\vc$ & $[\bc]_{\fl}$  &  $N^2 \times 1$ vector\\
 Current image estimate & $\bc_g$ &Updated each alg. step & $N \times N$ array\\
 Least squares solution & $\bc^\star$ & $\underset{\bc_g > 0}{\mathrm{argmin}}\ [Error(\bc_g)]$ & $N \times N$ array\\
 \hline
 \multicolumn{4}{|c|}{Point spread function (PSF)} \\
 \hline
  PSF   & $\psi_s(x,y)$    &  &  Real space \\
 Discrete PSF array &  $\bpsi_s$  &  $\psi_s(x_i,y_j)$   & $N \times N$ array\\
 Flattened PSF & $\PSFvec_s$ & $[\bpsi_s]_{\fl}$  &  $N^2 \times 1$ vector \\
 Concatenated set of PSFs & $\bpsi_T$ & $\left[\bpsi_1 | \bpsi_2|\bpsi_3|...|\bpsi_M\right]$ & $MN \times N$ array\\
 \hline
 \multicolumn{4}{|c|}{Optical transfer function (OTF)} \\
 \hline
 OTF   & $\Psi_s(k_x,k_y)$    & $\mathscr{F}\{\psi_s\} $ &  Complex space \\
 Discrete OTF array &  $\bPsi_s$  & $\DFT \, \bpsi_s$   & $N \times N$ array\\
 Flattened OTF & $\OTFvec_s$ & $[\bPsi_s]_{\fl}$  &  $1 \times N^2$ vector \\
 \hline
 \multicolumn{4}{|c|}{Noise} \\
 \hline
 LSM noise & $\bf n_s$ & Noise from detector  &  $N \times N$ array \\
 CSA noise & $\bf n_T$ & $\left[\bf n_1 | \bf n_2|\bf n_3|...|\bf n_M\right]$  &  $MN \times N$ array \\
 \hline
 \multicolumn{4}{|c|}{Data} \\
 \hline
 LSM data & $\bsig_s$ & $[\psi \circledast c]_{i,j}(x_i,y_j)+\bf n_s$  &  $N \times N$ array \\
 Flattened LSM data & $\vsig_s$ & $\left[\bsig_s\right]_{\fl}$  &  $N^2 \times 1$ vector \\
 CSA data & $\bsig_T$ & $\left[\bsig_1 | \bsig_2|\bsig_3|...|\bsig_M\right]$  &  $MN \times N$ array \\
 Flattened CSA data & $\vsig_T$ & $\left[\bsig_T\right]_{\fl}$  &  $ MN^2 \times 1 $ vector \\
 \hline
\end{tabular}

\begin{tabular}{ |p{3cm}||p{1.5cm}|p{3cm}|p{2.25cm}|  }
 \hline
 \multicolumn{4}{|c|}{\bf CSA super deconvolution variables cont'd} \\
 \hline
 Variable name & Notation & Construction & Domain or size \\
 \hline
 \multicolumn{4}{|c|}{Forward model} \\
 \hline
 LSM forward model vector& $\vec{\bf z}_s$ & $\mA_s \, \vc$  &  $N^2 \times 1$ vector \\
 LSM forward model array& $\bf z_s$ & $\bpsi_s \circledast \bc$  &  $N \times N$ array \\
 CSA forward model vector& $\vec{\bf z}_T$ & $\mA_T \, \vc$  &  $MN^2 \times 1$ vector \\
 CSA forward model array& $\bf z_T$ & $\bpsi_T \circledast \bpad(\bc)$  &  $MN \times N$ array \\
 \hline
 \multicolumn{4}{|c|}{Difference between forward model and data} \\
 \hline
 Error vector & $\vv_s$ & $\mA_s \, \left[\vc_g-\vsig_s \right]$  &  $N^2 \times 1$ vector \\
 Error array & $\bv_s$ & $\bpsi_s \circledast \bc_g-\bsig_s$  &  $N \times N$ array \\
 Concatenated error vector & $\vv_T$ & $\mA_T \, \left[ \vc_g-\vsig_T \right]$  &  $MN^2 \times 1$ vector \\
 Concatenated Error array & $\bv_T$ & $\bpsi_T \circledast \bpad(\bc_g)-\bsig_T$  &  $MN \times N$ array \\
 \hline
 \multicolumn{4}{|c|}{Regularization} \\
 \hline
 Regularization parameter & $\lambda$ & Constrains solution by limiting higher frequencies  &  Constant \\
  \hline
\end{tabular}

\section{Laser scanning image deconvolution}

In our specific case, we consider a set of ePSFs, with the s$^{\rm th}$ element denoted by $\psi_s(\pmb{\rho}_{\perp})$. Each PSF is the effective ePSF for laser scanning two photon microscopy, where the peak intensity of the illumination PSF is varied to provide spatial diversity in the ePSF through saturation of the excited state population. As the level of saturation is increased, the ePSF becomes wider, brighter, and steeper. The steeper ePSF implies higher spatial frequency content. The key to CSA is to combine all of this information to synthesize an estimated image of the specimen, which in our case is the spatial map of the fluorphore concentration $\bc$, from the set of images with the diverse saturated effective PSFs.




CSA relies on a set of well characterized saturated effective PSFs, $\psi_s$, that are used in combination with a forward model of the recorded signal. When considering continuous distributions, the convolution is given by the following integral
%
\begin{equation} \label{SIeq:conv}
    [\psi_s \circledast c](x,y)=\int_{-\infty}^{\infty}\int_{-\infty}^{\infty} \psi_s(x-x^{'},y-y^{'}) \, c(x,y) \,dx^{'} dy^{'}.
\end{equation}
As the data are acquired over a discrete array of scan positions $(x^{'},y^{'})$ and to enable computational estimation of the the specimen concentration on a discrete spatial grid, we consider the discreteized model of the convolution operator as a matrix $\mA_s$ that acts on a discrete object $\bc$. This two dimensional discrete convolution model can expressed as the matrix equation $\vsig_s =\mA_s \, \vc + \vn_s$. Measurement noise has been included as the noise vector $\vn$. For the sake of discussion, we will consider a set of $M$ measurements made with PSFs at varying levels of saturation, implying that the measurement index runs over the values $s=\{1, 2, \cdots, M \}$. The discrete approximation of the specimen is an image of $N \times N$ elements, so that the flattened object vector, $\vc$, has length $N^2$. As a result, each convolution operator matrix is $N^2 \times N^2$.

%

The convolution operator matrix, $\mA_s$, is constructed in the following way. Each shifted array, $\psi_s (x-x^{'},y-y^{'})$, is flattened (each column stacked on one another to create a vector) and placed into the rows of the convolution matrix. Multiplying this convolution matrix and the flattened second array, $\bc$, gives a vector which is the flattened two dimensional convolution of the two arrays. 
\begin{figure}[h!]
\centering\includegraphics[width=\linewidth]{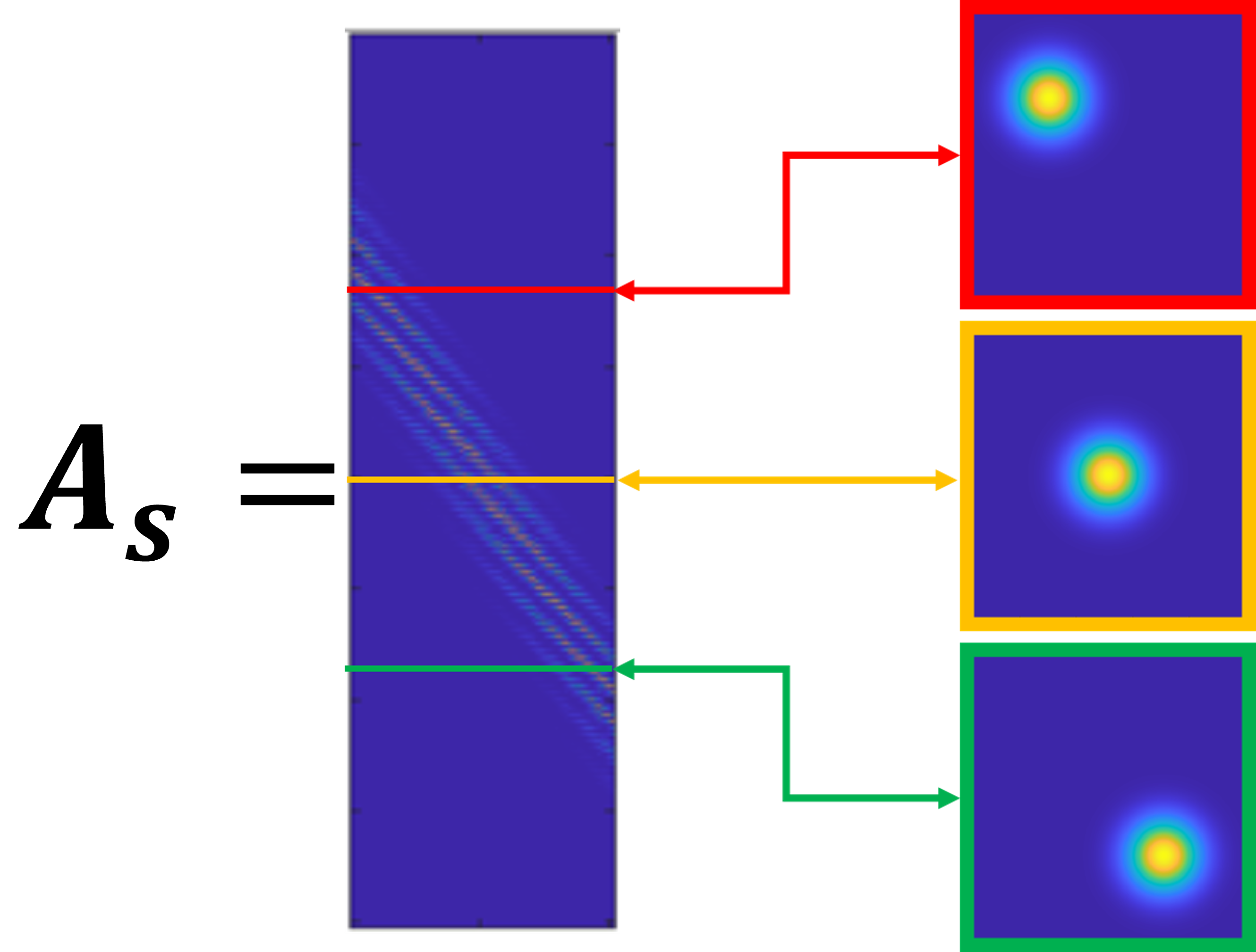}
\caption{Visual of the convolution matrix, $\mA$. Each row corresponds to a shifted point spread function $\bpsi_s$ which has been flattened.}
\label{fig:Amatfig}
\end{figure}

\begin{figure}[h!]
\centering\includegraphics[width=\linewidth]{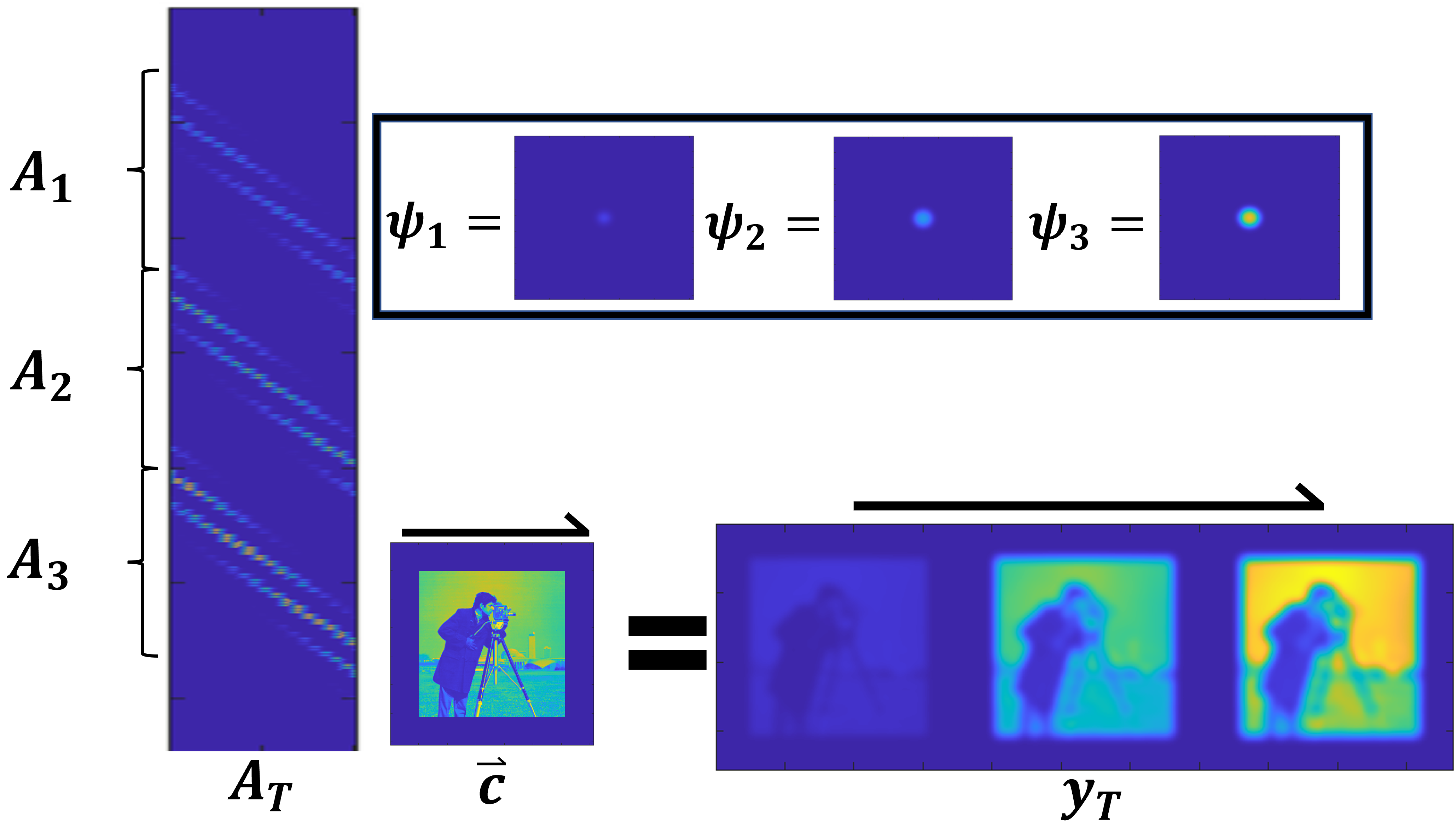}
\caption{Visual of the convolution matrix, $\mA_T$ operating on an object, $\bc$, with no noise}
\label{fig:A_T}
\end{figure}

To solve the super deconvolution problem posed by CSA, we consider the concatenated model shown in Fig. \ref{fig:A_T}. Here, each $\mA_s$ is stacked to produce the tall-skinny matrix $\mA_T$ that has a height of $MN^2$ rows and width of $N^2$. The concatenated convolution matrix, $\mA_T$, operates on the flattened object, $\vc$, to produce a concatenated signal, $\vsig_T$, of length $MN^2$, which is shown in an unflattened form, along with an unflattened object, in Fig.  \ref{fig:A_T}.

Now that we have established the concatenated signal as a linear matrix equation, $\vsig_T =\mA_T \, \vc + \vn_T$, in principle the solution is straightforward. Although each individual problem $\vsig_s =\mA_s \, \vc + \vn_s$ can be solved by applying a left acting inverse of $\mA_s$, with $\vc^\star = \mA^{-1}  \, \vsig_s$, this approach can not be applied to the concatenated data because $\mA_T$ is not a square matrix. In addition, it is well known that noise is a problem with the direct inverse. 




A least mean square (LMS) solution to the CSA problem can be obtained by using a left-acting Moore-Penrose pseudo-inverse that acts on the concatenated signal $\vsig_T$.  While this approach is not practicable for experimental data sets due to the data size and noise, we use the pseudo-inverse solution to characterize the expected performance of CSA and estimate the net PSF of the CSA super deconvolution that is optimal in a LMS sense. The CSA PSF is computed by generating noise-free data vectors for a set of simulated point spread functions that act on a simulated test object vector. The LMS image estimate obtained with CSA is then given by $\vec{\bc}^\star=\left(\mA_T^{\dagger}\mA_T+\lambda I \right)^{-1}\mA_T^{\dagger}\vsig_T$. Here we have considered a regularized pseudo-inverse where $\lambda$ is a constant called a regularization parameter which prevents small eigenvalues of $\mA_T^{\dagger}\mA_T$ from becoming very large after taking the inverse. This computation is feasible provided that the test object remains small. PSFs for CSA were computed with this strategy by defining the object as a narrow Gaussian spatial distribution with a width much smaller than the resolution of the net CSA super deconvolution. See the article for details.

The direct LMS solution not feasible for experimental data due to the large amount of computer memory consumed and infeasible computational times. The concatenated convolution matrix, $\mA_T$ is too large for most computers to be able to store in memory. For the case where the image size is $N \times N = 256 \times 256$, and accounting for zero padding, the matrix $\mA_T$ would take up 512GB of memory, which is much to large to carry out the computations mentioned above. Moreover, the pseudo-inverse matrix, which is $(M \times N^2)^2$ is completely impractical.


Efficient solution of the CSA super deconvolution problem then must resort to iterative methods that minimize a cost functional. Our implementation uses the cost functional given by 
\begin{equation}
    \vc^\star := \underset{\vc_g > 0}{\mathrm{argmin}} \frac{1}{2} \left(\norm{\mA_T \, \vc_g - \vsig_T}_2 + \norm{\lambda \,  I \, \vc_g}_2 \right).
    \label{SIeq:const}
\end{equation}
This cost functional is for the particular case with an error between the data and the estimated solution is given by the $L_2$ norm and Tikhonov regularization is applied. However numerical implementation of this optimization is still impractical if the entire matrix $\mA_T$ must be instantiated. 

Efficient solution of this problem is possible by avoiding the need to load $\mA_T$ into memory. The key to developing such an algorithm is to recognize that the majority of optimizer algorithms update the estimate of the object with a correction based on the scaled gradient of Eq. \ref{SIeq:const}. A simple calculation shows that this gradient is
\begin{equation}
    \vec{g_s}=\nabla_{\vc_g}\frac{1}{2} \left(\norm{\mA_T \, \vc_g - \vsig_T}_2 + \norm{\lambda \,  I \, \vc_g}_2 \right)=\mA_T^{\dagger} \left(\mA_T \, \vc_g - \vsig_T \right)+\lambda^2 \, I \, \vc_g.
    \label{SIeq:gradient}
\end{equation}
By making use of the Fourier convolution theorem, it is possible to efficiently evaluate the gradient given in Eq. \ref{SIeq:gradient} numerically without the need to use $\mA_T$ directly. Consider the convolution between one point spread function $\bpsi_s$, and our object $\bc$, as shown in Eq. \ref{SIeq:conv}. The objective is to use this expression and its adjoint to utilize the methods described above efficiently, but without the need of the convolution matrix $\mA_s$ directly. Finding an equivalent and more efficient computational strategy for performing the action of the $\mA_T$ and $\mA^\dagger_T$ enables efficient iterative calculations. 



\section{Efficient computation of a single deconvolution}

Before developing the efficient computational algorithm for the super CSA deconvolution, we will illustrate the approach with a conventional deconvolution. Consider the signal convolution given in Eq. \ref{SIeq:conv} written in terms of continuous forward, $\mathscr{F}\{\cdot\}$, and inverse  $\mathscr{F}^{-1}\{\cdot\}$, Fourier transforms 
\begin{equation}
    \psi_s \circledast c=\mathscr{F}^{-1}\{\mathscr{F}\{\psi_s\} \mathscr{F}\{c \}\}.
    \label{SIeq:singleconv}
\end{equation}
Eq. \ref{SIeq:singleconv} is equivalent to the matrix operation $\mA_s \, \vc$ and holds the key to understanding how to efficiently evaluate Eq. \ref{SIeq:gradient}. By treating the continuous Fourier transforms in Eq. \ref{SIeq:singleconv} as discrete Fourier transform (DFT) and discrete inverse Fourier transform (IFT) matrices denoted by $\DFT$ and $\iDFT$, respectively,  Eq. \ref{SIeq:singleconv} is given in matrix operator form as
\begin{equation} \label{convmat}
    \mA \, \vc=\iDFT \, \mathrm{\diag}\{ \OTFvec_s \} \, \DFT \, \vc.
\end{equation}
Here we identify $ \mA_s =\iDFT \, \diag\{ \OTFvec_s \} \, \DFT$. By comparison with Eq. \ref{SIeq:singleconv}, we obtain the computationally and memory efficient expression for $\mA \, \vc$ as
\begin{equation}
\label{SIeq:discreteAone}
    \mA \, \vc = \iDFT \left( \OTFvec_s \, \circ \, \DFT \, \bc \right)
\end{equation}
where for practical reasons the FFT and iFFT forms of the DFT an iDFT are used and the operations are based on Hadamard (element-wise) products, dented by $\circ$.  Here $\OTFvec_s = \DFT \, \PSFvec_s$ is the discrete OTF and $\PSFvec_s$ is the discrete PSF. When implementing Eq. \ref{SIeq:discreteAone}, the 2D $\DFT$ and $\iDFT$ are directly applied to the 2D PSF and object matrices, ${\bf \OTF} = \DFT \, {\bf \PSF}$ and $\bc$, respectively.

A similar formulation for the adjoint of the convolution operator, $\mA_s^\dagger$, is obtained by using the property $({\bf C \, D})^\dagger = \mathbf{D}^\dagger \, \mathbf{C}^\dagger $ and the fact that the DFT is a unitary operator, $\iDFT \DFT = \mathbf{I}$, which implies that $\DFT^\dagger = \DFT^{-1}$. Application of these properties produces the adjoint $ \mA_s^\dagger =\DFT \, \diag\{ \OTFvec_s^* \} \, \iDFT$. This expression shows that the DFT matrix $\DFT$, diagonalizes the convolution matrix $\mA_s$ with eigenvalues $\OTFvec_s$. When the adjoint acts on a vector, $\vv$, this is efficiently computed as
\begin{equation}
    \mA_s^\dagger \, \vv = \DFT \, ( \OTFvec_s^* \, \circ \, \iDFT \, \vv)
    \label{SIeq:matadj}
\end{equation}
The utility of this form of the adjoint operator becomes apparent if we identify $\vv$  with the quantity in parenthesis in Eq. \ref{SIeq:gradient}.

We gain insight into the nature of the adjoint operator by converting the matrix equation in Eq. \ref{SIeq:matadj} back into a continuous representation, which reads
\begin{equation} \label{convadj}
    \left(\psi_s \, \circledast  \right)^\dagger \,v =\psi_s \star v= \mathscr{F}^{-1}\{\mathscr{F}\{\psi_s\}^*\mathscr{F}\{v\}\}
\end{equation}
where $\star$ denotes the correlation operator. This observation is key to understanding how to modify the matrix operators used to construct the gradient term for the CSA super deconvolution.

\begin{algorithm2e}[H]
\DontPrintSemicolon
  
  $\bc_1=\mathrm{guess}$, $\alpha$ is the Lipschitz constant, $\bf prox()$ is the proximal operator which projects the current guess into the positive half space ($L_2$) or can be a shrinkage/thresholding operator ($L_1)$.
  
  \For{$k=1\xrightarrow[]{}N_{iter}$}
    {
        $\bc_k=\bc_k-\alpha\left[\left(\bpsi_s \star \left[\bpsi_s \circledast \bc_k-\bsig_s\right]\right)+\lambda^2 \bc_k\right]$
        
        $\bf x_k=prox(\bc_k)$
        
        $\bf t_{k+1}=\frac{1+\sqrt{1+4t_k^2}}{2}$
        
        $\bc_{k+1}=x_k+\frac{t_{k-1}}{t_{k+1}}(x_k-x_{k-1})$
    }
\caption{Deconvolution using FISTA}
\end{algorithm2e}






\section{Efficient computation of super CSA deconvolution}

Now, having a computationally efficient way of computing the gradient term in an optimization algorithm for the single image deconvolution problem using an efficient operator construction of $\mA_s$ and $\mA^\dagger_s$, we will now follow a similar strategy for the CSA super deconvolution. We now develop a super deconvolution form of the gradient operation in Eq. \ref{SIeq:gradient} that simultaneously makes use of the full data set. The super deconvolution requires that we establish modified forms of the discrete operator calculation for the forward operators $\mA_s$ and $\mA^\dagger_s$ in Eqs. \ref{SIeq:discreteAone} and \ref{SIeq:matadj}, that will accommodate the full concatenated data set, $\bsig_T$. Recall that the concatenated signal is composed of $M$ LSM images each of size $N\times N$, therefore having size $MN\times N$, where $\bsig_T = \left[ \, \bsig_1 \, | \,\bsig_2 \, \, | \cdots | \,\bsig_M \, \right]$, which is shown in Fig. \ref{fig:FFtxcorrCentral}. 

 To mimic the action of the concatenated set of convolution operators $\mA_T$ using Fourier transforms as in Eq. \ref{SIeq:singleconv} for the set of convolutions, the set of point spread functions $\bpsi_s$, are concatenated together into one large array in the same way as $\bsig_T$ to produce an array that contains $MN\times N$ elements, see Fig. \ref{fig:FFtxcorrCentral}. Unfortunately the discrete operator form for a single convolution given in Eq. \ref{SIeq:discreteAone} can not be directly applied when we replace $\bpsi_s$ with $\bpsi_T$ because the array $\bc$ is still only $N\times N$. This problem is easily remedied by padding $\bc$ with zeros in a way that makes it the same size as $\bpsi_T$ before the FFT operation, represented as $\bf pad()$ in equation \ref{convthmset}. This solves the dimensionality problem so that the Hadamard product can be performed, but also ensures the result of the operation of equation \ref{convthmset} yields an equivalent result as the output of the super convolution matrix $\mA_T$. This can be explained by considering that the construction of $\bpsi_T$ can be thought of as the sum of zero padded $\bpsi_s$'s that are the same size as $\bc$, but with the number of zeros on either side of the PSF dependent on the concatenation position. After application of the DFT operator, ($\DFT$), the resultant vector ${\bf \Psi}_T$ contains a sum of each OTF, but where each OTF has a linear phase ramp that corresponds to the offset in to recover the correct zero padding. The phase ramps are inherited through the Hadamard product, so that each PSF is present in the signal estimation operation.

\begin{equation} \label{convthmset}
    \mA_T\vc=\left[\bpsi_T \circledast \bpad(\bc)\right]_{\fl}
\end{equation}



\begin{figure}[h!]
\centering\includegraphics[width=\linewidth]{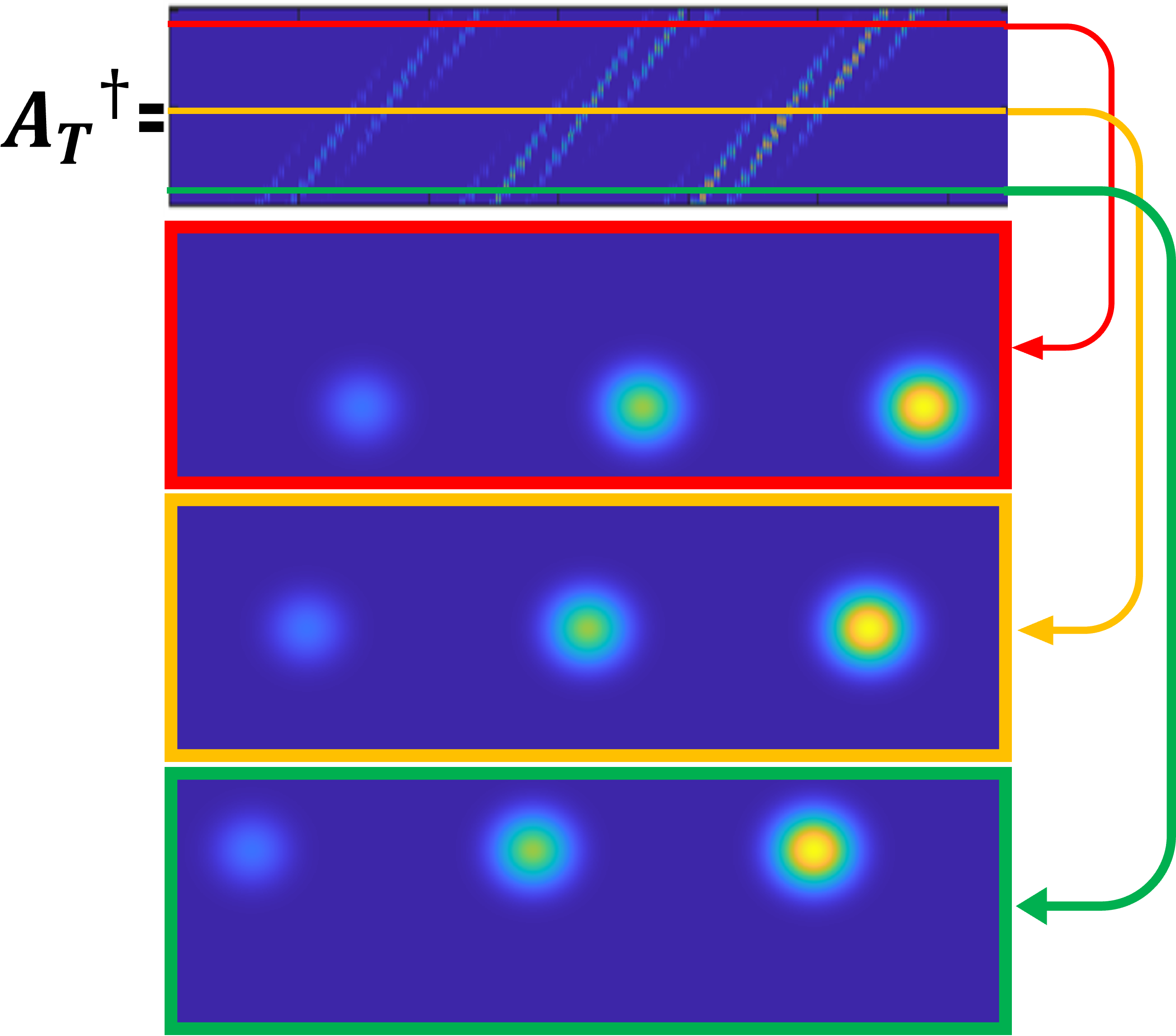}
\caption{Visual of the adjoint of the super convolution matrix, $\mA_T^\dagger$. Each row in this case corresponds to a shifted version of the concatenated set of point spread functions, $\psi_T$ which have been flattened into vectors. The maximum shift value is limited to the width of one of the concatenated point spread functions due to the limited number of columns contained in $\mA_T^\dagger$.}
\label{fig:AdaggOp}
\end{figure}

\begin{figure}[h!]
\centering\includegraphics[width=\linewidth]{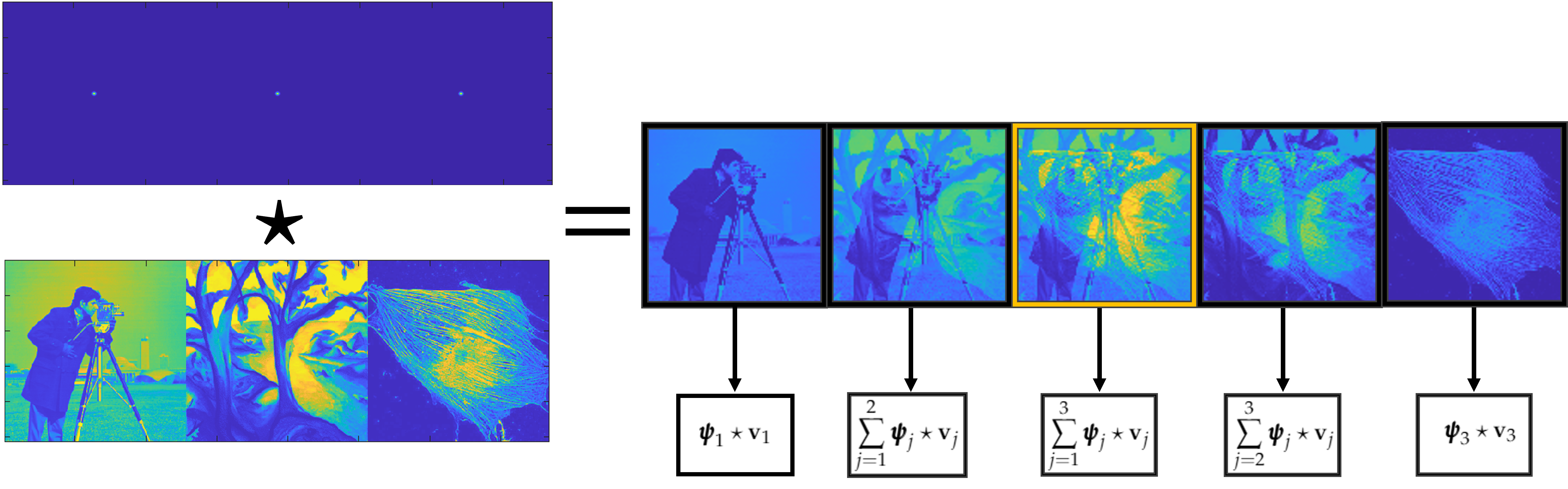}
\caption{Example of calculating the cross correlation between a set of three distinct images and a set of three delta functions using equation \ref{fullxcorr}. This figure demonstrates that the result of the cross-correlation between two arrays which are the concatenation of a set of images gives the concatenation of different sums of cross correlations between different combinations of images in the set. The image inside the yellow box is equivalent to $\mA_T^\dagger\vec{\bv_T}$ as described in Eq. \ref{CentralFFTxcorr}}
\label{fig:FFtxcorrCentral}
\end{figure}

The last step required in order to solve the super deconvolution problem is to find an equivalent expression as Eq. \ref{convadj} for the combined set of point spread functions, which will act as our $\mA_T^\dagger$ operation. With the updated $\vv_T$, we now need an efficient computation for the adjoint operation $\mA_T^\dagger \, \vv_T$. To avoid directly computing the full super adjoint operator, we leverage the observation that the adjoint operator is equivalent to a correlation (as shown in equation \ref{convadj}). Considering the discrete operator form of the adjoint given in Eq. \ref{SIeq:matadj}, we note that when we make the substitution $s \rightarrow T$, to transition from the single deconvolution to the super deconvolution, the result does not give the same answer as $\mA_T^\dagger\vv$, which is our goal. For an image, $\bc$, of size $N^2$ the matrix $\mA_T$ has width of $N^2$ and a height of $M \, N^2$. As a consequence, $\mA_T^\dagger$ takes in an input column vector of length $M \, N^2$ and gives an output column vector of length $N^2$, i.e., the image size, as shown in Fig. \ref{fig:AdaggOp}.



The output of $\bpsi_T \star \bv_T$ is $MN \times N$ (when flattened is a column vector of length $MN^2$). This discrepancy comes about because of the size of $\mA_T$; the number of columns are restricted so that it will only operate on an image of size $N \times N$. Consequently, this restricts the cross correlation carried out by $\mA_T^\dagger$ to the central part of the two concatenated arrays. The size of $\mA_T^\dagger$ is $N^2 \times MN^2$ so the maximum shift of $\bpsi_T$ with respect to $\bv$ when calculating the correlation is $N$. The cross correlation using Fourier transforms $\DFT^{-1}\{\DFT\{\bpsi_T\}^*\circ\DFT\{\bv_T\}\}$ gives the full cross correlation where the arrays have been shifted across each other by their full width $MN$. The result of $\mA_T^\dagger\bv$ is contained within $\bpsi_T \star \bv_T$ as the central part, of size $N \times N$, which can be retrieved by cropping. A nice corollary to this finding is that it demonstrates that the adjoint of the zero padding operation is cropping 

As shown in Fig. \ref{fig:FFtxcorrCentral}, the result of $\mA_T^\dagger \, {\bv_T}$ gives the sum of the cross correlations of each of the concatenated images in the set. This is because the correlation, similar to a convolution, records the overlap integral between the two arrays (with one now flipped) at different shifted positions which map to different points in the output. For the case of the operation of $\mA_T^\dagger$, each individual image within the concatenation is overlapped with the corresponding concatenated PSF array (i.e. each $\bpsi_s$ is lined up with the corresponding $\bv_s$). As the arrays are shift past one another, each individual image stays within its $s^{th}$ position due to the limited number of rows in $\mA_T^\dagger$.
\begin{equation} \label{Adaggcorr}
    \mA_T^\dagger\vec{\bv_T}=\left[\sum_{s=1}^M \bpsi_s \star \bv_s\right]_{\fl}
\end{equation}
Calculating the full cross correlation using $\DFT^{-1}\{\DFT\{\bpsi_T\}^*\circ\DFT\{\bv_T\}\}$ gives the concatenation of the cross correlation of combinations of elements in the set. In this case $\bpsi_T$, starts in a position where the $M^{th}$ image in $\bpsi_T$ overlaps with the first image in $\bv_T$ and is shifted until the first image in $\bpsi_T$ is overlapped with the $M^{\rm th}$ image in $\bv_T$ recording the overlap integrals along the way (Remembering that in a correlation, the array $\bpsi_T$ is flipped so that the image in the $M^{\rm th}$ position is actually $\bpsi_1$ when the correlation is being calculated). For example if $M=3$:
\begin{equation} \label{fullxcorr}
    \DFT^{-1}\{\DFT\{\bpsi_T\}^*\circ\DFT\{\bv_T\}\}=\left[\bpsi_1 \star \bv_1 \biggr\rvert \sum_{s=1}^2 \bpsi_s \star \bv_s \biggr\rvert \sum_{s=1}^3 \bpsi_s \star \bv_s \biggr\rvert \sum_{s=2}^3 \bpsi_s \star \bv_s \biggr\rvert \bpsi_3 \star \bv_3 \right] 
\end{equation}
We can calculate $\mA_T^\dagger \, {\bv_T}$ without the need for instantiating the full matrix $\mA$ into memory by taking the central part of the output of $\DFT^{-1}\{\DFT\{\bpsi_T\}^*\circ\DFT\{\bv_T\}\}$. This operation is equivalent to $\mA_T^\dagger\vec{\bv_T}$.
\begin{equation} \label{CentralFFTxcorr}
    \mA_T^\dagger\vec{\bv_T}=\left[C \left(\DFT^{-1}\{\DFT\{\bpsi_T\}^*\circ\DFT\{\bv_T\}\}\right)\right]_{\fl}
\end{equation}
The operator $C$ crops out the central image from the cross correlation.

We now have the ability to calculate the operation of both $\mA_T$ and $\mA_T^\dagger$ without needing to hold the super convolution matrix in memory and we can perform this operation in a computationally efficient manner by using the Fast Fourier Transform (FFT). Using the techniques above for the super deconvolution, the gradient of the error function which we have defined previously can be calculated using equation \ref{JointGradfft}.
\begin{equation} \label{JointGradfft}
    \nabla_{\vc_g}\frac{1}{2} \left(\norm{\mA_T\vc_g - \vsig_T}_2 + \norm{\lambda \,  I \, \vc_g}_2 \right)=\left[C\left(\bpsi_T \star\left(\bpsi_T \circledast \bpad(\bc_g) - \bsig_T \right)\right)\right]_{\fl}+\lambda^2I\vc_g
\end{equation}
This technique combines information contained in a set of images which each have been convolved with a unique point spread function, with the algorithm pseudocode given in Algo. \ref{algo:superdeconv}. Combining this information together to form a single, high resolution, deconvolved image performs better than traditional single deconvolution.  

\begin{algorithm2e}[H]
\DontPrintSemicolon
  $\bc_1=\mathrm{guess}$, $\alpha$ is the Lipschitz constant, $\bf prox()$ is the proximal operator which projects the current guess into the positive half space ($L_2$) or can be a shrinkage/thresholding operator ($L_1)$.
  
  \For{$k=1\xrightarrow[]{}N_{iter}$}
    {
        $\bc_k=\bc_k-\alpha\left[C\left(\bpsi_T \star \left[\bpsi_T \circledast \bpad(\bc_k)-\bsig_T\right]\right)+\lambda^2 \, \bc_k\right]$
        
        $\bf x_k=prox(\bc_k)$
        
        $\bf t_{k+1}=\frac{1+\sqrt{1+4t_k^2}}{2}$
        
        $\bc_{k+1}=x_k+\frac{t_{k-1}}{t_{k+1}} \, (x_k-x_{k-1})$
    }

\caption{Super deconvolution using FISTA}
\label{algo:superdeconv}
\end{algorithm2e}

\section{Data and PSF normalization}

\begin{figure}[h!]
\centering\includegraphics[width=\linewidth]{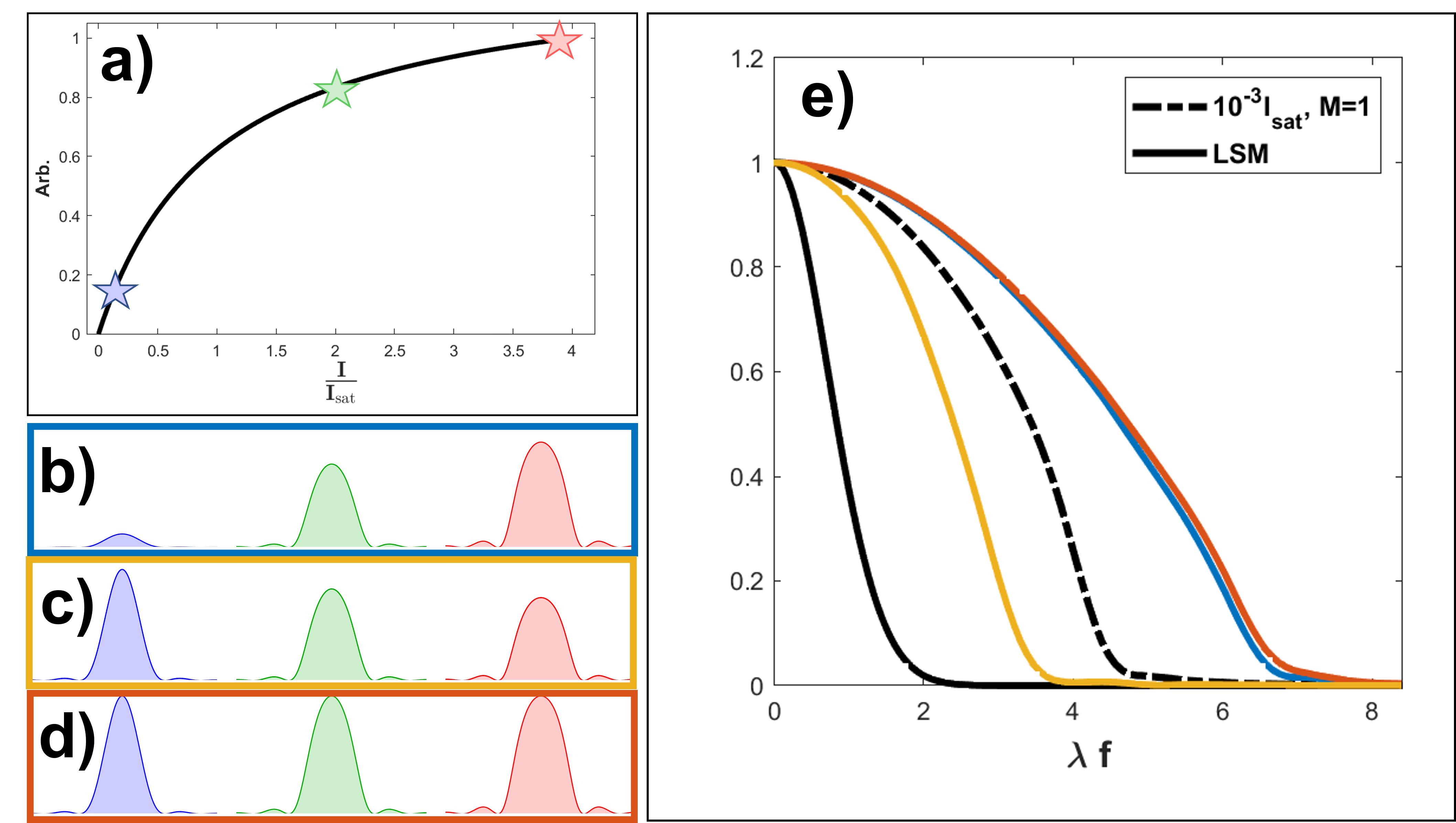}
\caption{Plot e) shows the differences in imaging performance by re-scaling the PSFs in the set $\bpsi_T$ in different ways by comparison of their simulated MTFs using the pseudoinverse (M=15 for each of the colored curves going up to $4I_{sat}$). Plot a) shows the measured two photon saturation curve along with a fit of the data. Plots b)-d) show line outs of the simulated PSFs for different saturation levels being normalized in different ways. The Plots in b) show how the PSfs are scaled according to the saturation curve, c) shows them normalized in terms of energy and d) they are normalized to the peak value. This shows the maximal imaging performance is gained when normalizing the PSFs to their peak value.}
\label{fig:PSFnorm}
\end{figure}

One method to obtain a diverse set of point spread functions for super deconvolution imaging (SDI), which is the method that we demonstrate here, is to drive the excitation of molecules that produce fluorescent emission at increasing excitation intensity. As the excitation intensity increases, the average excited state population becomes saturated, which  produces saturated effective point spread functions (ePSFs). Examples of some of the saturated ePSFs are shown in Figs. \ref{fig:PSFnorm} and \ref{fig:AsymMTFs}. As the illumination intensity, and thus the peak saturation parameter, increases, the saturated ePSF becomes broader and exhibits steeper edges. The steeper edge at high saturation levels also pushes the OTF to a larger range of spatial frequency support as seen in  Fig. \ref{fig:AsymMTFs}. The set of saturated ePSFs adds spatial frequency diversity to the image data, from which stable super resolution images are extracted. Moreover, overlap in the spatial frequency support across the set of saturation levels drives stability in the estimated object because the information in these spatial frequency ranges must be self-consistent in the estimated image. The set of images each blurred with varied ePSFs, each at a particular saturation level, is used in the SDI super resolution image reconstruction. The result is improved SNR across a much broader range of spatial frequency values -- producing a significantly higher resolution image than one obtains by deconvolving a single image in the set as is shown in Fig. \ref{fig:PSFnorm} (e). CSA also outperforms the related super resolution fluorescent imaging techniques called saturated excitation (SAX) microscopy as we shown in Fig. (3) in the accompanying paper.

The total emitted intensity of fluorescent power follows the average excited state population values. When the excitation intensity exceeds the saturation intensity, the ground-state state population becomes, on average, depleted, producing a saturation in the excited state population. The effect of the saturation of the excited state population is detected in the saturation of emitted fluorescent power with increased excitation intensity and is displayed in a saturation curve. Fig. \ref{fig:PSFnorm} (a) shows an experimental saturation curve measured for the fluorphore used in the main manuscript. The saturated ePSFs for several saturation levels, indicated by the stars, are shown in Figs. \ref{fig:PSFnorm} (b)-(d).  Each of these sub-panels highlights are particular normalization of the ePSFs. Because the average fluorescent power increases with the level of saturation, the effective PSF brightness also increases accordingly. Images recorded at higher saturation levels increase in brightness by a factor of the ratio of the saturation intensities. Fig. \ref{fig:PSFnorm} (b) is normalized in terms of saturation level, and thus in terms of the emitted fluorescent power. In contrast, Fig. \ref{fig:PSFnorm} (c) displays the ePSFs normalized by power (normalized with respect to area, i.e., $2\pi \int \mathrm{ePSF}(\rho) \, \rho \, d\rho$) and Fig. \ref{fig:PSFnorm} (d) shows normalization with respect to the peak ePSF intensity. 

A set of recorded images will display a total signal level (or brightness) that is proportional to the saturation level, which is indicated by the ePSF normalization in Fig. \ref{fig:PSFnorm} (b). Given the variation in image, and the effective PSF, brightness with a change in the saturation of the illumination, the question of how to normalize the data for the super deconvolution algorithm naturally arises. More specifically, how should the image data and the modelled ePSFs be normalized relative to each other? Clearly, at each saturation level, the ePSF and the image data must be scaled in the same way to avoid model mismatch, but does weighting the ePSFs in different ways affect the imaging performance? 

Three normalization strategies are immediately evident: 1) use the collected image power and the ePSF are scale with the level of saturation (Fig. \ref{fig:PSFnorm} b), 2) normalize each image and ePSF to have the same energy (Fig. \ref{fig:PSFnorm} c), and 2) normalize each image and ePSF to have the peak value in the spatial domain (Fig. \ref{fig:PSFnorm} d). To determine which normalized method produces the optimal estimated object spatial frequency distribution, we evaluate each method in the context of a least mean squared solution to the point image. The super deconvolution MTF for each of these normalization strategies are shown in Fig. \ref{fig:PSFnorm} (e), where the color of the MTF matches the color of the box for each normalization strategy. These simulated MTFs are found by reconstructing a point-like object, that is much smaller than the resultant PSF, with the pseudoinverse of noise-free simulated data. The MTF support for the LSM in a weak excitation, and thus approximately linear, regime and the deconvolution of a single image are shown in the solid black and dashed black lines in Fig. \ref{fig:PSFnorm} (e) to enable a direct comparison between normalization strategies, as well as the MTFs presented in the paper.



As discussed previously, using the pseudoinverse requires the ability to use a very large amount of memory. These calculations were performed on a computing cluster at Colorado State University (CSU) with 192GB of RAM. To solve the  deconvolution problem directly by computing a pseudoinverse requires that the problem be scaled down in order to be able to run successfully. Each image is set to a size $125 \times 125$, and the image field of view is set to a low enough range to enable a large enough spatial frequency support when estimating the OTF. As can be seen in figure \ref{fig:PSFnorm} the best imaging performance is achieved when the ePSFs and data are all normalized to their peak value, but this is only marginally better than a scaling strategy based on the saturation level as is obtained directly from an experiment.

\section{Asymptotic behavior of imaging performance with degree of saturation}

The SDI technique combines information from a set of LSM images that exploits diversified spatial frequencies from a set of ePSFs that each have a spatial frequency diversity. In particular, in CSA that we demonstrate here, we inject spatial frequency diversity into our ePSFs by driving the fluorescent excitation of molecules into saturation. As mentioned previously, higher levels of saturation produce ePSFs that are both spatially broadened and steeper at the edges, which leading to a sampling of higher spatial frequency values at higher levels of saturation. Here, we address the question of how this broadened spatial frequency support behaves as we drive the excitation deep into saturation by increasing the peak excitation intensity. This result is remarkable given the limitations of single point scanning saturated excitation deconvolution \cite{Laporte:14}.

The question of the asymptotic behavior of CSA at increasingly high saturation levels is addressed by computing the expected PSF/OTF distributions as the total peak saturation intensity is increased to large values.  Computed MTFs plotted with increasing saturation parameter are shown in Fig. \ref{fig:AsymMTFs} (a). The predicted cutoff frequency of the CSA technique follows the logarithmic curve with the saturation level shown in Fig. \ref{fig:AsymMTFs} (b). This curve shows that above a certain level of saturation the rate of improvement slows. The expected resolution improvement rises very quickly up until about $100 \, I_{sat}$.



Clearly, high levels of saturation will bring an excellent return in the increase of the span of spatial frequency support, which will bring about a commensurate improvement in the spatial resolution of SDI. Of course, an important question to address is whether we can make use of high levels of saturation for a realistic experimental situation. For continuous wave (cw) laser excitation, the saturation intensity for a 3-level molecular system, where stimulated absorption is negligible due to internal conversion, is given by \cite{Harke:08} $I_{\rm sat} = h \, \nu_i / \tau_e \, \sigma_{\rm abs}$. Here, $h$ is Planck's constant, $\nu_i = c/\lambda_i$ is the optical frequency of the illumination light, $c$ is the vacuum phase velocity of electromagnetic radiation, $\lambda_i$ is the illumination light, $\tau_e$ is the excited state lifetime, and $\sigma_{\rm abs}$ is the absorption cross section. There are a wide range of saturation levels depending on the spectroscopic properties. Of particular interest for saturation-driven super resolution microscopy are luminescent transitions with a metastable upper level state.

Many applications for super resolution imaging are biological in nature, so we will take some values of the constants above that are typical for an organic dye molecule: $\lambda_i = 500$ nm, $\sigma_{\rm abs} = 3 \times 10^{-16}$ cm$^2$, and $\tau_e = 4$ ns, which gives a saturation intensity of $I_{\rm sat} \approx 330 $ kW/cm$^2$ \cite{Harke:08}. Other luminescent probes used for a broad range of applications with a wide range of saturation intensities based on variations in the absorption cross section and excited state lifetime. Some of the most popular probes include fluorescent nanodiamonds (FNDs)  \cite{Laporte:16}, $I_{\rm sat} \approx 1.3$ MW/cm$^2$; rare earth dopants \cite{REnumbers,ZhongGoldner+2019+2003+2015}, $I_{\rm sat} \approx 70$ kW/cm$^2$; quantum dots \cite{QDnumbers, doi:10.1063/1.4800445}, $I_{\rm sat} \approx 21$ kW/cm$^2$, and photoswitchable proteins \cite{PSPreview, Zhang10364}, $I_{\rm sat} \approx 1-100$ W/cm$^2$.

The limit of the peak saturation parameter $\alpha_0 = I_0/I_{\rm sat}$ that could be used depends on many factors such as damage inflicted to the sample through ablation, ionization, or heating, or by destruction of the probe through photobleaching. Clearly, the peak saturation value is further dependent on the saturation intensity, and probes with a very low saturation intensity could be imaged with extremely high spatial resolution. To appreciate the range of possible saturation parameters, we set an upper bound by considering the order of magnitude for the intensity required for ionization of materials of $I_{\rm ion} \sim 10^{15}$ W/cm$^2$. While ablation or other damage mechanisms may kick in a lower intensities, some material systems can withstand intensities right up to this level. This upper bound corresponds to a peak saturation level of $\sim 8 \times 10^8$ for FNDs, which have the highest saturation intensity in our list above. Inspection of Fig. \ref{fig:AsymMTFs} indicates that we can achieve significant improvements in imaging resolution orders of magnitude below the ionization and damage limits. Stimulated emission depletion (STED) microscopy is routinely used with high intensities in the depletion beam, with values typically exceeding $I_{\rm STED} \sim 1$ GW/cm$^2$ when organic dyes are used \cite{Harke:08}. At this saturation intensity, we reach a peak saturation parameter of $\alpha_0 \sim 3000$ for a typical organic dye and over $10^5$ for both rare earth dopant probes and photoswitchable proteins. While the actual value of the peak saturation parameter that can be used in any particular experiment depends on details, these numbers indicate that the values shown in Fig.  \ref{fig:AsymMTFs} could be accessed under the appropriate circumstances. One other note with regard to the STED comparison that should be made is that in the case of STED, the high intensity depletion beam switches the molecules to a dark state, which suppresses photobleaching. Thus, the high average excitation in the case of CSA could force the use of lower intensities under conditions with pervasive photobleaching.

The asymptotic behavior of CSA shows the potential to attain exceptional imaging resolution. Note that the parameters in Fig.  \ref{fig:AsymMTFs} are scaled based on wavelength, and they are calculated for a numerical aperture of $\mathrm{NA} = 1.4$. In these plots, we observe an increase in spatial frequency support up to $\lambda \, f_c/2\mathrm{NA} \sim 5$, or an imaging resolution down to $\delta \rho \sim 35.7$ nm for an excitation wavelength of $\lambda_i = 500$ nm (using diffraction limited resolution of $f_c=\frac{2NA}{\lambda}$). This scaling is robust even when considering noise in the measurement, because higher saturation levels produce a higher SNR signal.


\begin{figure}[h!]
\centering\includegraphics[width=\linewidth]{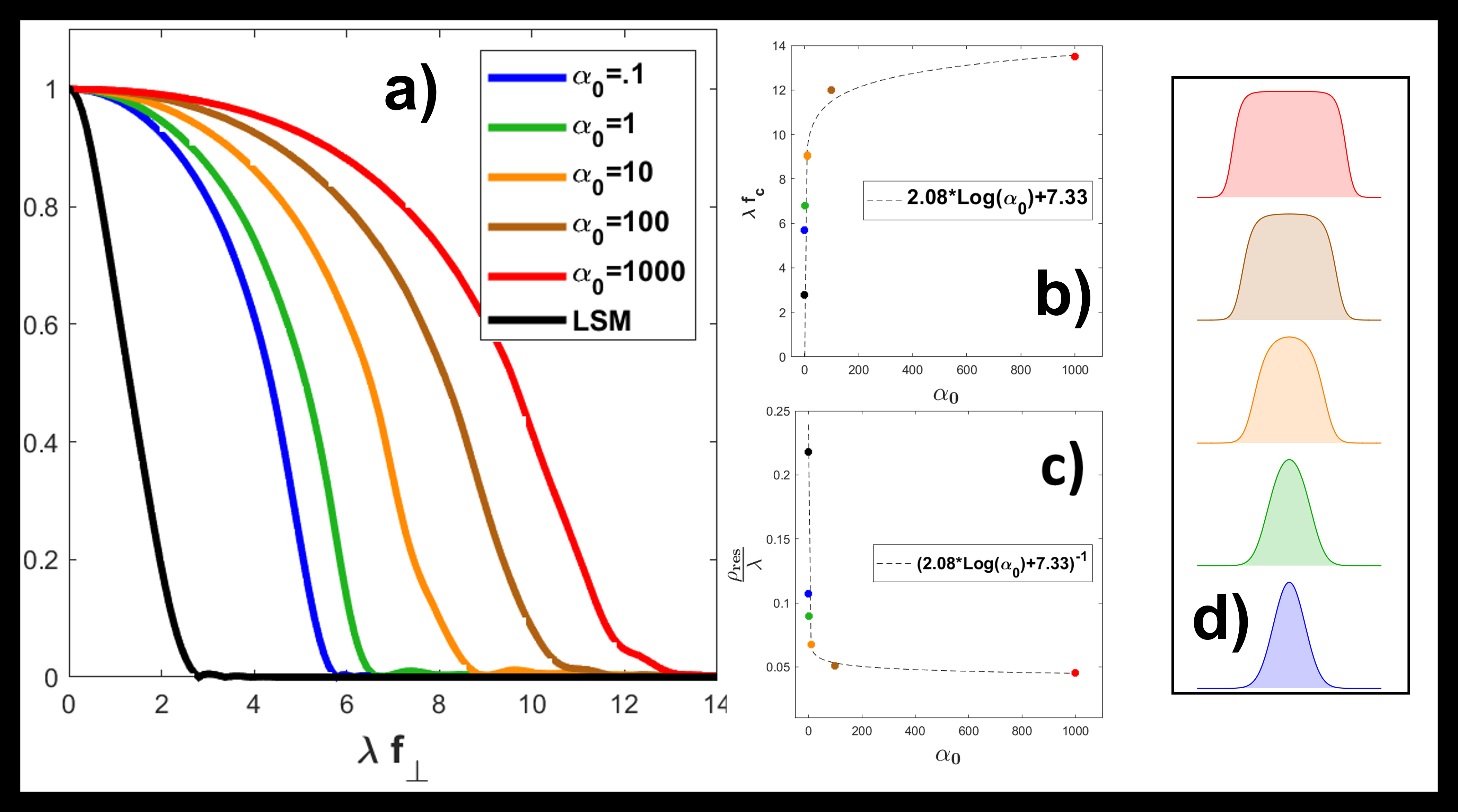}
\caption{Plot a) shows the expected MTF for increasing levels of saturation in the asymptotic limit. Plots b) and c) show how the expected resolution trends with max saturation level. Plots a), b), and c) are normalized in terms of wavelength so they are all unitless. Panel d) shows how the PSF changes shape according to the maximum saturation level of each simulation. Each MTF is generated from using the CSA technique with a set of fifteen PSFs evenly sampling the saturation curve starting from $.01 \, I_{sat}$ to the maximum saturation level. The plots to the right show the PSFs at increasing levels of peak saturation ($\alpha_0 = I_0 /I_{sat}=[.1, 1, 10, 100, 1000]$ respectively)}
\label{fig:AsymMTFs}
\end{figure}

\section{Measurement of the saturation curve}

In order to implement the CSA technique, a model of the change in ePSF with increased excitation intensity is required. As the ground state population depletes, the excited state population follows a nonlinear excitation curve with respect to the excitation intensity. 
While this relationship can be predicted, the saturation intensity may not be well known a priori, and depends on many experimental factors such as the sample environment. Moreover,   the functional dependence on the saturated emission intensity curve depends on the laser pulse duration, the pulse shape, the absorption nonlinearity, and the laser repetition rate.
To sidestep building a complex model, we
experimentally measuring the saturation curve of the sample being imaged.

To measure the saturation curve of a sample, the excitation beam of the microscope is parked at a particular location. The power of the illumination beam directed into the microscope is rapidly swept by varying the control signal to an acousto-optic modulator (AOM). A portion of the emitted fluorescent emission is detected on a photo-multiplier tube (PMT) after passing through a filter which rejects the pump light. The AOM can be rapidly scanned to avoid sample damage and photobleaching.  Another advantage to using an AOM to modulate the input intensity is many cycles of the ramp function can be recorded rapidly and averaged to reduce the error in the final reported saturation curve. 

\begin{figure}[h!]
\centering\includegraphics[width=\linewidth]{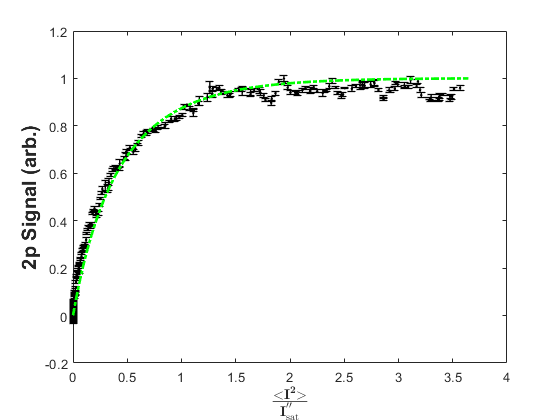}
\caption{Measured saturation curve for two-photon excitation of fluorescein dyed fibers. Black error bars indicate measured points of the saturation curve averaged over 10 cycles. The error is taken as the standard deviation of the set of measurements of each point. The green dashed line shows a fit of the data using equation \ref{eq:TPFmeancount} for pulsed two-photon excitation. Fitting to the data using this equation is useful for estimating the saturation intensity, which gives an idea how far above saturation we have reached ($I_{\rm sat}^{(2)}$ is the two-photon saturation squared intensity and $I$ is the input intensity)}
\label{fig:2psatcurve}
\end{figure}

Power levels are measured at the peak and minimum of the ramp to be able form the relationship between input power and corresponding saturation level. Having measured the saturation curve, 
the ePSFs at a given saturation level can be simulated with a measured unsaturated PSF. This  unsaturated PSF is obtained by recording an image of an object that is much smaller than the resolution of the system, such as a fluorescent nano-diamond (FND). The ePSF is then simulated by feeding the unsaturated PSF in the the nonlinear saturation curve, so that the ePSF amplitude is distorted by the local saturation level across the spatial distribution of the illumination light intensity, with a distortion that is scaled by the peak saturation value at the peak of the PSF. In CSA imaging a set of LSM images are taken of the sample at increasing intensity values that correspond to increasing peak saturation parameters. The input power is measured for each image in the set so that this can be compared to the measured saturation curve in order to model the matching ePSF for each image. Figure \ref{fig:2psatcurve} shows a measurement of the saturation curve of two-photon excitation of fluorescein dyed fibers.

In order to estimate the saturation level attained from the measured saturation curve, a fit to the data is made using an equation predicting the saturation curve where the only free parameter is the saturation intensity. In this case Equation \ref{eq:TPFmeancount} is used which shows the mean fluorescence count, $F$, for two photon emission assuming that the pulse duration and the spacing of the pulses in the laser pulse train are much shorter than the fluorescence lifetime. The square of the two-photon saturation intensity is given by $I_{\rm sat}^{(2)} = 2 \, h \, \nu_i \, (\tau_t \, \sigma_{\rm TPA})$
and $<I^2>$ is average squared intensity of the input \cite{DavisSat}. Here $h$ is Planck's constant, $\nu_i$ is the illumination optical frequency, $\tau_e$ is excited stat (e.g., fluorescence) lifetime, and $ \sigma_{\rm TPA})$ is the two photon absorption cross section. The emitted two-photon fluorescent signal follows
\begin{equation}
    F \propto \frac{1-e^{-\frac{2<I^2(t)>}{I_{\rm sat}^{(2)}}}}{2+\left(1-e^{-\frac{2<I^2>}{I_{\rm sat}^{(2)}}}\right)}.
    \label{eq:TPFmeancount}
\end{equation}

\begin{figure}[h!]
\centering\includegraphics[width=\linewidth]{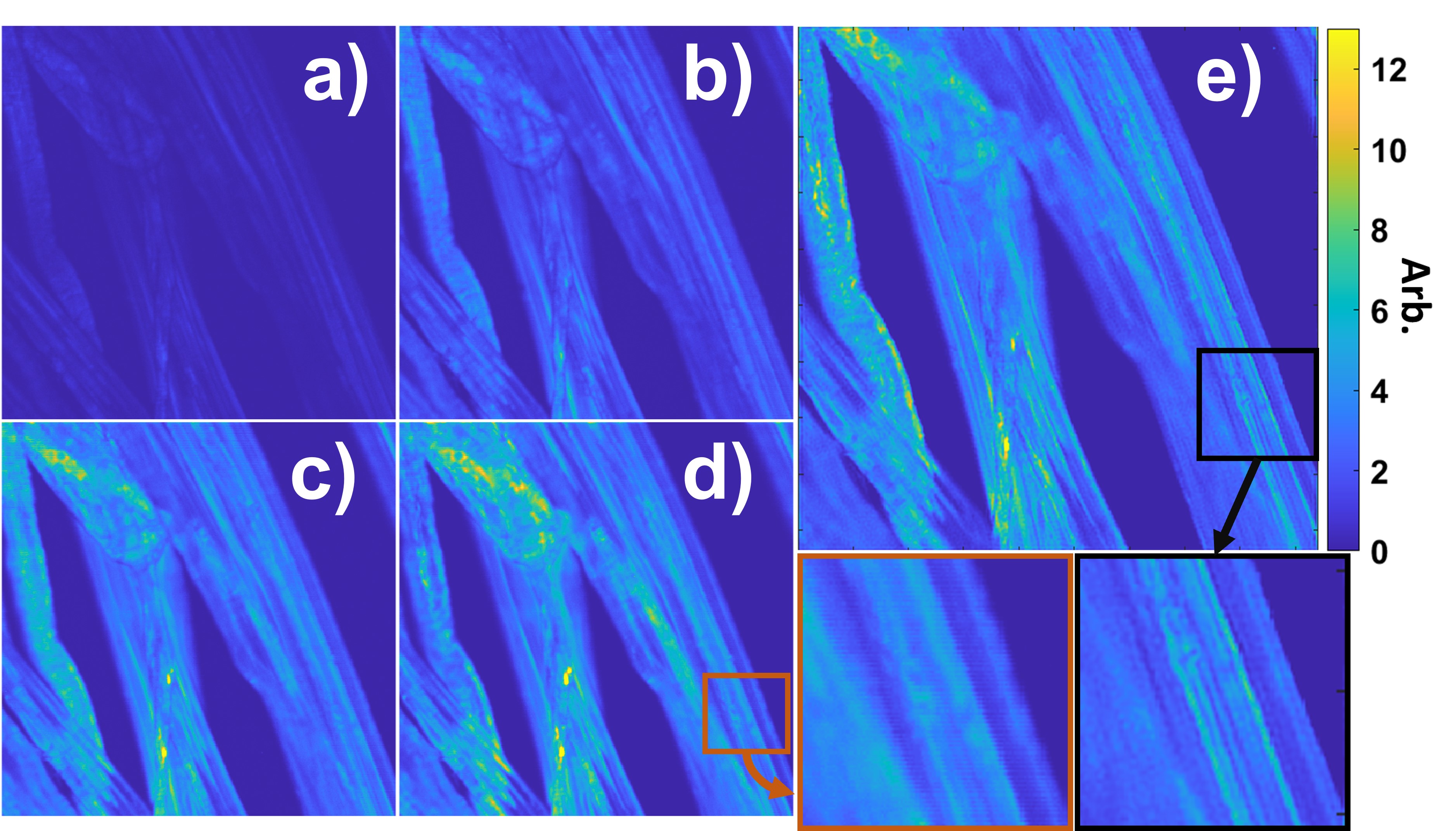}
\caption{Example point-scanned images taken at different saturation levels (a-d) contained in the set used for the CSA reconstruction e). The scale bar to the right of e) applies to all images shown.}
\label{fig:ExampleImages}
\end{figure}


\section{Multi-Photon SPIFI}

\begin{figure}[h!]
\centering\includegraphics[width=\linewidth]{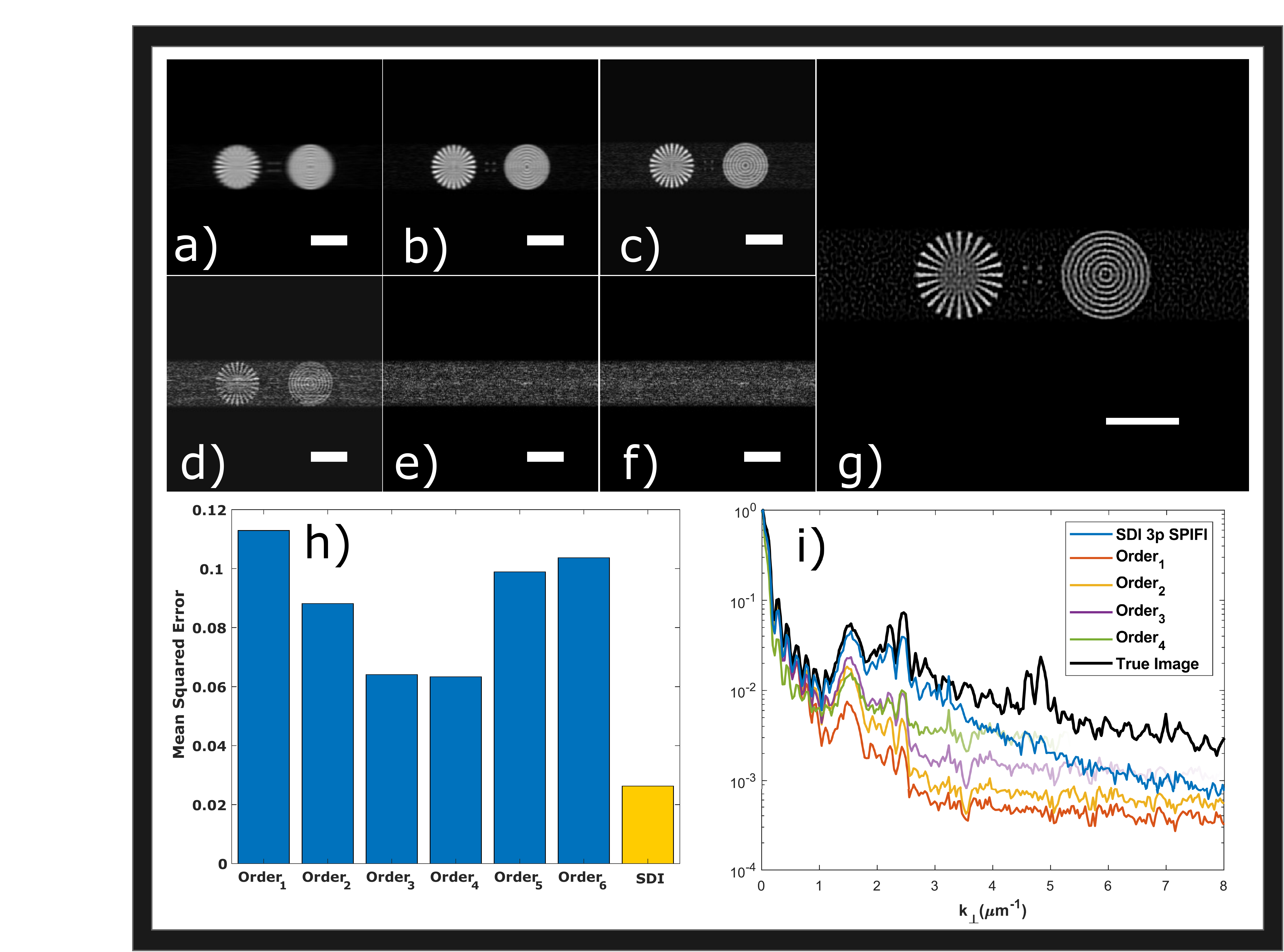}
\caption{Images a-f show the MP-SPIFI images for orders 1-6 respectively. Image g) shows the reconstructed image jointly deconvolving the six images MP-SPIFI images with their modeled point spread functions. The bar graph shown in h) shows the mean squared error of the six images compared to the reconstructed one using SDI. The plots in i) show the radial averages of the spatial frequency content in all of the corresponding images compared to that of the ground truth image. }
\label{fig:MPSPIFIReconstructions}
\end{figure}

The SDI algorithm was tested on another nonlinear single-pixel imaging method called multi-photon SPIFI (MP-SPIFI) in which a line focus is modulated with a linearly swept illumination spatial frequency and then the signal power is collected with a single pixel detector. \cite{Futia:11, Schlup2011, Higley:12, Hoover2012, Howard2013, Winters:15, Field:15, Field6605, Field:16, Worts:18, Field:18, STOCKTON201824, Stockton2019, Heuke:20} The simulation makes use of a Poisson noise model \cite{HARWIT197944} applied to the expected photon signal count for three photon absorption fluorescent emission MP-SPIFI microscopy. Harmonic orders were extracted form each image order (Fig. \ref{fig:MPSPIFIReconstructions}) which were then jointly deconvolved using SDI and imaging transfer functions for SPIFI. \cite{Stockton:20, STOCKTON2022128401}

This shows the generality of the SDI method as it can be applied to many different types of imaging modalities.

\nocite{*}

\bibliography{SDI}

\end{document}